\documentclass{article}
\usepackage[utf8]{inputenc}
\usepackage[left=1cm, right=1cm, top=2cm, bottom=2cm]{geometry}
\usepackage{changepage}
\usepackage[superscript,biblabel]{cite}
\usepackage{hyperref}
\usepackage{booktabs}
\usepackage{caption}
\usepackage{bm}
\usepackage{amsmath}
\usepackage{listings}
\usepackage{array}
\usepackage{soul}
\usepackage{amssymb}
\usepackage{changepage}
\usepackage{bbold}
\usepackage{natbib}

\usepackage{fancyvrb}

\usepackage{epsfig,amsfonts,graphicx,url}
\usepackage{mathtools,eucal}
\usepackage{xr}
\usepackage{color}

\DefineVerbatimEnvironment{Sinput}{Verbatim}{fontshape=sl}
\DefineVerbatimEnvironment{Soutput}{Verbatim}{}
\DefineVerbatimEnvironment{Scode}{Verbatim}{fontshape=sl}

\DefineVerbatimEnvironment{Code}{Verbatim}{}
\DefineVerbatimEnvironment{CodeInput}{Verbatim}{fontshape=sl}
\DefineVerbatimEnvironment{CodeOutput}{Verbatim}{}
\newenvironment{CodeChunk}{}{}

\newcommand{\pkg}[1]{{\fontseries{b}\selectfont #1}}
\let\proglang=\textsf
\let\code=\textsf

\title{\textbf{Joint Modeling of Multivariate Longitudinal and Survival Outcomes with the \proglang{R} package \pkg{INLAjoint}}}

\author{\textbf{Denis Rustand} \ \ \
        \textbf{Janet van Niekerk}\ \ \
        \textbf{Elias Teixeira Krainski}\ \ \
        \textbf{H\aa vard Rue}\\ Statistics Program, Computer, Electrical and Mathematical Sciences and Engineering Division \\ King Abdullah University of Science and Technology \\ Thuwal 23955-6900, Kingdom of Saudi Arabia}

\date{}

\begin{document}

\maketitle

\abstract{\begin{adjustwidth}{40pt}{40pt}
\normalsize This paper introduces the \proglang{R} package \pkg{INLAjoint}, designed as a toolbox for fitting a diverse range of regression models addressing both longitudinal and survival outcomes. \pkg{INLAjoint} relies on the computational efficiency of the integrated nested Laplace approximations methodology, an efficient alternative to Markov chain Monte Carlo for Bayesian inference, ensuring both speed and accuracy in parameter estimation and uncertainty quantification. The package facilitates the construction of complex joint models by treating individual regression models as building blocks, which can be assembled to address specific research questions. Joint models are relevant in biomedical studies where the collection of longitudinal markers alongside censored survival times is common. They have gained significant interest in recent literature, demonstrating the ability to rectify biases present in separate modeling approaches such as informative censoring by a survival event or confusion bias due to population heterogeneity. We provide a comprehensive overview of the joint modeling framework embedded in \pkg{INLAjoint} with illustrative examples. Through these examples, we demonstrate the practical utility of \pkg{INLAjoint} in handling complex data scenarios encountered in biomedical research.\ \\\end{adjustwidth} \begin{adjustwidth}{20pt}{120pt}
$^*$\textbf{Correspondence:} Denis Rustand, KAUST, Email: denis@rustand.fr\end{adjustwidth}}

\normalsize 
\section[Introduction]{Introduction}
%% Note: If there is markup in \(sub)section, then it has to be escape as above.
Longitudinal and survival data analyses have received important attention in statistical literature in recent decades, the simultaneous consideration of these data types in the context of joint models presents a unique challenge and opportunity. Indeed, when examining repeated measurements of a biomarker alongside an event of interest, an inherent association between these two outcomes often exists. The risk of the event can be influenced by the longitudinal biomarker, with biomarker measurements typically being truncated by the occurrence of the event. Joint models, integrating longitudinal and survival components, have become indispensable for capturing the intricate interplay between time-varying endogenous variables and survival outcomes. Examples include the analysis of longitudinal CD4 lymphocytes counts and AIDS survival \citep{wulfsohn1997joint}, prostate-specific antigen dynamics and cancer recurrence \citep{Proust09}, patient reported outcomes and progression free survival in cancer clinical trials \citep{Hatfield12}, cancer tumor dynamics and mortality risk \citep{Rustand20}, longitudinal markers related to graft function and survival in patients with chronic kidney disease that received renal transplantation \citep{Gould15}, aortic gradient dynamics and cardiac surgery outcomes \citep{Andrinopoulou14}, amongst many more.

Endogenous longitudinal markers are variables whose values or future trajectory directly correlates with event status. Using time-dependent Cox models is inadequate for endogenous markers due to their inability to appropriately handle the inherent characteristics of these variables, such as the direct relationship with event status, the susceptibility to measurement error and the missingness process due to discontinuous measurement times. Joint modeling emerges as a more suitable and accurate alternative in such cases, facilitating a comprehensive understanding of the complex interplay between covariates and event outcomes. They enhance statistical inference efficiency by simultaneously utilizing both longitudinal biomarker measurements and survival times. Fitting multiple regression models simultaneously and accounting for a possibly time-dependent association is computationally challenging. Two-stage approaches were initially proposed \citep{de1994modelling, tsiatis1995modeling}, involving fitting the longitudinal and survival regression models separately and combining the fitted models in a second stage. However, these models suffer from important limitations, either because they overlook informative drop-out due to survival events or because they require to fit many longitudinal models (i.e., at each event time), leading to an infeasible computational burden and unrealistic assumptions on the distribution of random effects \citep{wulfsohn1997joint}. The simultaneous fit of the multiple components of a joint model was introduced a couple years later \citep{faucett1996simultaneously, henderson2000joint}, mitigating some bias observed with two-stage approaches and enhancing efficiency of statistical analyses by leveraging information from both data sources simultaneously, although long computation times were necessary for these types of models.

Recent advances in computing resources and statistical software have enabled the estimation of various joint models, but despite the interest in analyzing more than a single longitudinal outcome and a single survival outcome concurrently, most existing inference techniques and statistical software have limitations on such comprehensive joint models because of the computational challenge as stated in the literature \citep{Huang12, hickey2016joint, li2021flexible}. Indeed, common estimation strategies such as Newton-Raphson, expectation-maximization or MCMC (Markov chain Monte Carlo) face limitations in scalability and convergence speed, particularly for complex data and models (i.e., many outcomes or parameters). In addressing this concern, the \proglang{R} package \pkg{INLAjoint}, introduced in this paper, emerges as a solution designed to address the need for a reliable and efficient estimation strategy for joint models. With a flexible architecture, \pkg{INLAjoint} allows statisticians and researchers to construct intricate regression models, treating univariate regression models as modular building blocks that can be assembled to effortlessly form complex joint models. \pkg{INLAjoint} leverages the power of the integrated nested Laplace approximations (INLA) methodology, which emerged as an efficient alternative to MCMC methods for Bayesian inference of latent Gaussian models \citep{Rue09}. INLA capitalizes on the sparse representations of high dimensional matrices to provide rapid approximations of exact inference. It has been recognized as a fast and reliable alternative to MCMC for Bayesian inference of joint models, with the capability to handle the increasing complexity inherent in joint models \citep{Rustand23, rustand2023bis}. The standard joint model often employs a shared combination of fixed and random effects to analyze longitudinal outcomes and associated survival outcomes. An alternative is the latent class joint model \citep{Proust14}, which assumes the population comprises homogeneous groups with similar biomarker trajectories and event risks, which is out of the scope of \pkg{INLAjoint}.

While many methods implemented in different software were proposed to fit joint models with a single longitudinal and a single survival outcome, the available software for multivariate joint models is more limited. Here, we focus on methods that can (at least in theory, in practice many of these methods often fail to converge for joint models) deal with multiple longitudinal and/or multiple survival outcomes. In \proglang{SAS} \citep{SAS03}, the procedure \pkg{NLMIXED} has been widely used to fit various joint models \citep{Guo04, Zhang14} but it has poor scaling properties due to the curse of dimensionality of the Gauss-Hermite quadrature method used to do an anylitical approximation of the integral over the random effects density in the likelihood of joint models. This multivariate integral is the main reason why joint models are computationally expensive. In \proglang{Stata}, the \pkg{merlin} command has been proposed to provide a unified framework able to fit various joint models \citep{Crowther20} but it suffers from convergence problems and long computation time \citep{Olivares23}. In \proglang{R} \citep{R23}, many packages have been introduced to fit various joint models, the most well-known being \pkg{JM} \citep{Rizopoulos10} which is limited to a single longitudinal outcome but can accomodate competing risks of event, a pseudo-adaptive Gaussian quadrature is used to integrate out random effects in the likelihood, which has better scaling properties compared to the standard quadrature but remains limited. The \proglang{R} package \pkg{joineRML} \citep{Hickey18} uses a Monte Carlo expectation-maximization algorithm and can include multiple longitudinal outcomes but is limited to Gaussian distributions for the longitudinal markers and a single survival outcome. The \proglang{R} package \pkg{frailtypack} \citep{Rondeau12} can handle the simultaneous inclusion of recurrent events modeled with a shared frailty survival model and a terminal event along with a single longitudinal outcome using the Levenberg-Marquardt algorithm (i.e., a Newton-like algorithm). 

While the aforementioned softwares are based on a frequentist framework, Bayesian inference has gained a lot of interest recently in the context of joint models. Although it is philosophically different, both approaches can be used for similar purposes and can be compared under some conditions. Indeed, the ``maximum a posteriori'' provided under Bayesian inference is equal to the ``maximum likelihood'' when the prior distributions plays a negligible role (i.e., non-informative prior), since the posterior is then driven mainly by the data and matches the maximum likelihood estimates. Simulation studies \citep{Rustand23, rustand2023bis} for joint models show that frequentist statistics computed over joint models fitted with Bayesian inference, can be as good if not better compared to those obtained with frequentist inference (i.e., when data is not informative for the value of a parameter, frequentist inference will fail while Bayesian inference will return information from the prior, i.e., explicitly showing that the data is not informative for this parameter). Bayesian inference usually rely on sampling-based algorithms such as MCMC, resulting in a significant computational burden. The \proglang{R} package \pkg{JMbayes} \citep{Rizopoulos16} is the ``Bayesian version'' of \pkg{JM} using MCMC to fit joint models, it can handle multiple longitudinal outcomes of different types and also multiple survival submodels such as competing risks or multi-state. It is not fully Bayesian as it relies on a corrected two-stage approach. More recently, the \proglang{R} package \pkg{JMbayes2} \citep{JMbayes2} has been proposed using a full Bayesian approach with MCMC, it can also handle multiple longitudinal outcomes of different types, competing risks and multi-state models as well as frailty models for recurrent events. It has better scaling properties compared to \pkg{JMbayes} as it uses parallel computations over chains and an efficient MCMC implementation in \proglang{C++}. Finally, the \proglang{R} package \pkg{rstanarm} \citep{Brilleman20} can deal with up to three longitudinal outcomes of different nature (at the moment) but is limited to a single survival outcome and it relies on the Hamiltonian Monte Carlo algorithm as implemented in \proglang{Stan} \citep{Carpenter17}, which has slow convergence properties for joint models \citep{Rustand23}.

All these softwares, while capable of fitting joint models, cannot be pictured as direct competitors but rather as benchmarks against which the flexibility of \pkg{INLAjoint} can be highlighted. Comparisons of some of these approaches with the INLA methodology have already been proposed in the literature. \cite{rustand2023bis} compared the \proglang{R} package \pkg{INLA} and \pkg{frailtypack} in simulation studies to fit a joint model for a semicontinuous longitudinal outcome and a terminal event in the context of cancer clinical trial evaluation and found that \pkg{INLA} was superior to \pkg{frailtypack} in terms of computation time and precision of the fixed effects estimation. The frequentist estimation faced some limitations and led to convergence issues when fitting a complex joint model. \cite{Rustand23} compared \pkg{INLAjoint} with \pkg{joineRML}, \pkg{JMbayes2} and \pkg{rstanarm} in multiple simulation studies involving up to three longitudinal markers along with a terminal event. It showed that INLA is reliable and faster than alternative estimation strategies, has very good inference properties and no convergence issues. However, while the \proglang{R} package \pkg{INLA} that implements the INLA methodology has been previously introduced to fit joint models \citep{Niekerk21}, the increasing complexity related to the inclusion of multivariate outcomes makes it cumbersome to use this package directly and motivated the development of the \proglang{R} package \pkg{INLAjoint} as a user-friendly implementation specifically for joint models. As an illustration, the joint model fitted in the application of \cite{Rustand23}, including 7 regression models for 5 longitudinal markers and 2 competing risks of event, was initially fitted with \pkg{INLA} and required more than 1000 lines of code while fitting the exact same model with \pkg{INLAjoint} requires 15 lines of code. This new package makes the use of INLA for joint modeling more user friendly and widely applicable, it avoids some commonly made mistakes in the code that can have important implications. \pkg{INLAjoint} uses a more friendly syntax than the \proglang{R} package \pkg{INLA} both for fitting joint models and in the output summary, moreover it offers more extensive tools appropriate for joint models compared to plain \pkg{INLA} such as specific plots and predictions as highlighted in this paper. In this context, this paper aims to underscore the versatility of \pkg{INLAjoint} in constructing models that are currently unattainable with other software alternatives.

The paper is structured as follows: we first introduce the methodological framework in Section \ref{metsec} with details on the range of models that can be fitted with the \proglang{R} package \pkg{INLAjoint} and the underlying methodology of INLA. We then describe the package structure and usage in Section \ref{pkgIJ} and provide usage examples in Section \ref{UseEx} before concluding with a discussion.

\section[Methodological framework]{Methodological framework}
\label{metsec}
In this section, we first introduce and briefly outline the specifics of the INLA methodology. While algorithms like MCMC are more general and capable of handling a broader spectrum of models, INLA can only be applied to models that can be formulated as LGMs (latent Gaussian models). However, the flexibility of LGMs comes to the forefront, as many diverse models, including mixed effects, temporal, spatial, and survival models, can be effectively formulated within the LGM framework. This characteristic makes INLA a powerful tool for a wide range of practical applications, showcasing its versatility in accommodating various models that fall under the umbrella of LGMs. We then present the methodology employed by \pkg{INLAjoint} to conceptualize different regression models for both longitudinal and survival data as LGMs. This formulation establishes a cohesive framework, aligning these models with the INLA methodology. Consequently, it enables the creation of versatile joint models that encompass both longitudinal and survival components. Finally, this section concludes with a brief overview of the INLA method to compute the posterior distributions of interest and its recent enhancements, with relevant references for more details.

\subsection{Unified regression framework based on latent Gaussian model structure}
LGMs represent a specific subset of hierarchical Bayesian additive models. It is essential to note that the INLA methodology is tailored precisely for this family of models, offering computational efficiency in estimating model parameters and the associated uncertainty. In particular, the inherent conditional independence within the hierarchical structure of LGMs leads to sparse precision matrices, a key characteristic that INLA exploits. This strategic approach allows INLA to achieve not only fast but also highly accurate inference of the posterior distribution of the parameters within the LGM framework, establishing its effectiveness in handling complex models efficiently.

Let $\bm u$ denote the vector of parameters included in a regression model that can be specific for an individual or a group of observations (i.e., fixed and random effects). In the context of Bayesian inference, these parameters are associated to a distribution that leads to an additional set of parameters $\bm \omega$, referred to as hyperparameters (e.g., variance of the fixed effects, variance-covariance of random effects or likelihood parameters related to the data distribution).

An LGM is defined with a specific hierarchical structure, with three layers:
\begin{enumerate} 
\item The first layer is the likelihood of the observed data, $\bm D$, conditional on the latent field and the hyperparameters $p(\bm D| \bm u, \bm \omega)$. There is no restriction for the distribution of the observed data and any reasonable likelihood can be used. An important feature here is that the likelihood contribution of each data point $D_i$ is independent conditional on the latent field and the hyperparameters, therefore the full likelihood is given by the product of the likelihood for each data point, the latent Gaussian field and the hyperparameters priors such that the marginal likelihood is defined as
$p(\bm D) = \prod_i^N p(\bm D_i| \bm u_i, \bm \omega) p(\bm u_i | \bm \omega) p(\bm \omega)$, where N is the number of observed data points.
\item The second layer is the latent field, $\bm u$, with  prior $p(\bm u | \bm \omega)$. These parameters are constrained to be Gaussian, leading to a multivariate Gaussian distribution of the latent field, often referred to as the latent Gaussian field. This is the core of latent Gaussian models and because of this constraint, INLA can take advantage of mathematical tools that rely on this feature. It is important to note here that models with non-Gaussian random effects cannot fit with the LGM framework and cannot be fitted with INLA. However, while other estimation strategies like MCMC allow for non-Gaussian random effects, all the packages mentioned in the introduction are also limited to Gaussian distributions for random effects at the moment.
\item The last layer is the prior knowledge on the hyperparameters distribution $p(\bm \omega)$. Similar to the likelihood, there is no restriction on the prior for the hyperparameters, i.e., they can have any distribution (some options are already implemented and user-defined distributions can be added)
\end{enumerate}

INLA is particularly efficient when the multivariate Gaussian distribution of the latent field has a sparse precision matrix, indicating that some elements in the latent field are independent of many others when conditioned on a few. Such sparsity characterizes a Gaussian Markov Random Field (GMRF), as detailed in \citet{Rue05}. INLA capitalize on the sparse structure of precision matrices in LGMs, using modern numerical algorithms for sparse matrices to achieve efficient Bayesian inference.

\subsection{Regression models expressed as LGMs}
\label{LGMreg}
We first define a general framework for regression models, let $\eta_{ijk}$ denote the linear predictor for individual $i (i=1, ..., N_k)$ at time $j (j=1, ..., N_{ik})$ for outcome $k (k=1, ..., K)$. This predictor is a linear combination of fixed effects $\bm \beta_k$ associated to covariates $\bm X_{ijk}$ and random effects $\bm b_{ik}$ associated to covariates $\bm Z_{ijk}$. The relationship between the observed data and the linear predictor is defined as follows:
$$g(\mathrm{E}[D_{ijk}]) = \eta_{ijk} = \bm X_{ijk}^\top \bm \beta_k + \bm Z_{ijk}^\top \bm b_{ik},$$
where $g(\cdot)$ is a link function. It is easy to picture how this model formulation is part of the LGM framework as the fixed and random effects are part of the latent Gaussian field $\bm u$ while the parameters related to their distribution are part of the hyperparameters vector $\bm \omega$. While it is trivial to match this framework with mixed effects models (where multiple longitudinal outcomes can be linked through the correlation of their random effects), it is important to detail how survival models and joint models for longitudinal and survival data fit with this framework.

\subsubsection{Survival regression models as LGMs}
The most common approach to fit survival data assume proportional hazards, as introduced by \cite{Cox72}. A few years later, it has been shown that the Cox model can be approximated with a Poisson regression \citep{Holford76, Whitehead80, Johansen83}. The key idea is to convert the continuous-time hazard function into a discrete-time event count. If the time intervals are small enough, the hazard can be considered roughly constant within each interval. Then, the number of events in each interval follows a Poisson distribution, where the mean is proportional to the hazard in that interval.

Let $T_i$ be the survival time of individual $i$, and  $\bm X_i(t)$ be the covariates vector, possibly time-dependent. The hazard function for the Cox model is given by:

$$h\left(t|\bm X_i(t)\right) = h_0(t) \cdot \exp\left(\bm \beta ^\top \bm X_i(t)\right),$$

where $h_0(t)$ denotes the baseline hazard, and $\bm \beta$ a vector of coefficients for covariates $\bm X_i(t)$.

The Poisson approximation in discrete time is based on the assumption that the number of events $N_i$ in a small time interval of length $\Delta t$ follows a Poisson distribution:

$$N_i \sim Poisson(\lambda_i)$$

where the mean $\lambda_i$ is proportional to the hazard within the interval such that:

$$\lambda_i = h\left(t_i|\bm X_i(t)\right) \Delta t.$$

This approximation becomes more accurate as the time intervals become shorter. However, it's important to note that this approximation may introduce bias if the hazard changes rapidly within the intervals but if each unique death time is made into its own interval the Cox model can be replicated exactly. In practice, we use an offset term to match exactly the event time and censoring time when they occur within an interval. Moreover, since more intervals involve a more flexible approximation of the baseline, \pkg{INLAjoint} uses random walks instead of piecewise constant approximation to avoid overfitting and ensure a smooth baseline hazard function. The Poisson regression model then fits with the framework described in section \ref{LGMreg} and can be fitted with INLA. Parametric proportional hazards models as well as accelerated failure time models can also be part of this framework as they simply involve a specific distribution for the approximation of the baseline hazard (e.g., exponential, Weibull). Right, left or interval censoring are common censoring schemes and can be accommodated in our approach. The contribution to the likelihood can then easily be defined for censored and observed event times, as done with the Cox model.

\subsubsection{Joint models as LGMs}
Joint modeling involves fitting multiple longitudinal and/or survival outcomes simultaneously. When fitting multiple longitudinal outcomes, the outcome-specific mixed effects regression models can be linked through the correlation of their random effects. On the other hand, when fitting longitudinal and survival outcomes or multiple survival outcomes, the models can be linked by sharing a linear combination of fixed and random effects from a submodel with another one. In both cases, the contribution to the likelihood of each submodel is independent conditional on these correlated or shared effects. The likelihood of a joint model then corresponds to the integral over random effects density of the product of each submodel's likelihood where a survival model's likelihood is defined as the product of individual likelihoods while a longitudinal model's likelihood is the product of individual likelihoods across measurement occasions.

The main challenge in fitting joint models is the multivariate integral over the density of random effects in the likelihood function, which is usually handled by numerical approximation such as Monte Carlo sampling or Gauss quadrature methods which are both time-consuming. 
With INLA both the fixed and random effects are treated jointly as a latent Gaussian field (LGF) as the fixed effects are given independent Gaussian prior distributions with zero mean and fixed variance while the Gaussian prior for each random effect are specified using sparse precision matrices, which gives a joint sparse precision matrix for the LGF (all the fixed and random effects). The first Laplace approximation uses a Gaussian approximation for the distribution of the LGF conditional on the hyperparameters in place of the likelihood integration which speed up the method as only few inner iterations involving a sparse linear solver are needed allowing the high dimentional LGFs (we have fitted models with 2 millions random effects).

Moreover, a second challenge is the association structure that links submodels for different outcomes. While correlated random effects are time invariant, it is common to share time-dependent components such as the individual deviation from the population mean as defined by random effects (shared random effects parametrization), the entire linear predictor (current value or current level parametrization) or the derivative over time of the linear predictor (current slope parametrization). These time-dependent associations complexify the contribution to the likelihood of survival models as the usual approach involves an analytical solution for the contribution to the likelihood of survival models because the likelihood contribution of observed and censored events involve the survival of the individual up to the event or censoring time. The survival function is based on the cumulative risk (i.e., the integral of the risk function), when a time-dependent component is included in the risk function this integral needs to take into account the evolution of the time-dependent component during follow-up to compute the contribution to the likelihood. While other software deal with this additional integral with sampling or numerical approximation methods such as Gaussian quadrature, INLA takes advantage of the decomposition of the follow-up into small intervals to account for the evolution of time-dependent components in the risk function as described in the previous section, which again avoids the need for numerical approximation techniques and fits within the LGM framework. This completely avoids the cumbersome numerical approximation and can be arbitrarily accurate by using arbitrary small intervals. In the context of the use of parametric functions for the baseline risk, there is no need for such decomposition of the follow-up. However, nothing prevents from using the same data augmentation strategy as for the semi-parametric approach, by decomposing the follow-up into small intervals. The use of left truncation and right censoring allows to reconstruct the entire follow-up until the event or censoring time of an individual which adds up to the full contribution to the likelihood of the individual and can accomodate time-varying covariates. This is also particularly useful for prediction purpose, where we are able to know the risk at any time point conditional on the time-dependent components. This method is illustrated in the first usage example of this paper. While it sounds time-consuming to increase the data size for model fitting, the INLA method is specifically designed to take advantage of latent Gaussian models structure and the cost in speed of this technique is minimal compared to the use of numerical approximation of integrals. Overall, integrals in the likelihood are handled as conditional probabilities directly resulting from Bayes theorem. More details on the formulation of joint models as LGMs is available in \citet{Martino11}, \citet{Niekerk19} and \citet{Niekerk21}.  

\subsection{Bayesian inference of LGMs with INLA}

The central challenge lies in approximating the posterior distributions of the parameters of the LGMs, where the latent field has a multivariate Gaussian prior distribution, and the precision matrix depends on the model's hyperparameters. The natural sparse structure of the precision matrix in LGMs forms the core of INLA's efficiency, leveraging state-of-the-art numerical algorithms for sparse matrices.

There has been a lot of enhancement of the INLA methodology since it was first proposed by \citet{Rue09} and described in the context of fitting joint models in \citet{Niekerk21}. These improvements make the INLA methodology uniformly faster while keeping it as accurate as the original formulation. A detailed description of the ``modern'' INLA algorithm is proposed in \citet{Niekerk23} and this section highlights these modifications while giving a brief overview of INLA. 

The first step of the INLA methodology consists in an approximation of the marginal posterior distribution of the hyperparameters given the observed data $p(\bm \omega | \bm D)$ using Laplace method (i.e., integrate out the fixed and random effects through efficient sparse matrix computations to reduce the optimization problem to only the hyperparameters in order to find their posterior mode). Recently, the ``smart gradient'' approach has been implemented in order to more efficiently find the mode of hyperparameters in this first step \citep{Fattah22}.

In the second step of INLA, the conditional posterior distributions of each element of the latent field $u_i$ are approximated, given the hyperparameters and the observed data $p(u_i|\bm \omega, \bm D)$ with Laplace approximation. However, this second nested Laplace approximation for the fixed and random effects, can be replaced by an implicit (low rank) variational Bayes correction for the already available Gaussian approximation from the first Laplace approximation \citep{Niekerk21b}, which gives the same accuracy as the Laplace approximation at a much lower computational cost, as illustrated in \cite{Niekerk23}. The third step uses a numerical integration to integrate out the hyperparameters and obtain the marginal posterior distribution of $u_i$:
$$p(u_i|\pmb{D})\approx \sum_{h=1}^{H}\tilde{p}(u_i|\bm\omega_h^*,\pmb{D})\tilde{p}(\bm\omega^*_h|\pmb{D})\Delta_h,$$
where the integration points $\bm \omega^*_1, ..., \bm \omega^*_H$ are defined based on the density of the hyperparameters $\bm \omega_h$ associated to weights $\bm \Delta_h$. For this integration part over the hyperparameters, the original formulation of INLA was limited to a small number of hyperparameters (i.e., 20) but the modern formulation of INLA has been further optimized recently and we successfully applied \pkg{INLAjoint} to fit models with more than 150 hyperparameters.

While in the classic INLA formulation, the latent field includes the linear predictors in addition to its components (fixed and random effects) for computational purposes as described in \citet{Rue09}, it is not the case anymore in the modern INLA formulation, where it is now possible to infer the linear predictor posteriors through a linear combination of the fixed and random effects, post-inference. This contributes to the increased speed by reducing the size of the matrices involved in computations. Indeed, the size of the latent field does not depend on the size of the data anymore, thus enabling the use of the INLA methodology for large data applications.

\section[The R package INLAjoint]{The R package INLAjoint}
\label{pkgIJ}
In order to use \pkg{INLAjoint}, one first needs to install the INLA algorithm implemented in the \proglang{R} package \pkg{INLA} with the following command:
\begin{CodeInput}
R> install.packages("INLA",repos=c(getOption("repos"),
+  INLA="https://inla.r-inla-download.org/R/stable"), dep=TRUE)
\end{CodeInput}
More informations on the INLA installation can be found on the website: \\
https://www.r-inla.org/download-install \\
The \proglang{R} package \pkg{INLAjoint} is available from CRAN at https://CRAN.R-project.org/package=INLAjoint. It can be installed in \proglang{R} with the command:
\begin{CodeInput}
R> install.packages("INLAjoint")
\end{CodeInput}
The development version of the package is available on GitHub and can be installed using \pkg{devtools} \citep{Wickham22} with the following command:
\begin{CodeInput}
R> devtools::install_github("DenisRustand/INLAjoint")
\end{CodeInput}
The \proglang{R} package \pkg{INLAjoint} is a comprehensive tool for fitting various joint models, it features a primary model-fitting function named `joint()`:
\begin{CodeInput}
joint(
  formLong = NULL, dataLong = NULL, id = NULL, timeVar = NULL,
  family = "gaussian", link = "default", corLong = FALSE,
  corRE = TRUE, formSurv = NULL, dataSurv = NULL, 
  basRisk = "rw1", NbasRisk = 15, cutpoints = NULL, 
  assocSurv = NULL, assoc = NULL, control = list(), ...
)
\end{CodeInput}
This function returns instances of an S3 class, compatible with relevant S3 methods such as \code{summary}, \code{plot}, and \code{predict}. We first describe the function's parameters related to the modeling of longitudinal outcomes through mixed effects regression models. Then we introduce parameters specific to survival models and joint modeling before concluding the section with details about relevant S3 methods.

\subsection{Longitudinal submodels}
We start by describing the available models for longitudinal data, which are defined as mixed effects models. Several likelihoods have been implemented, allowing to choose the appropriate model given the specific distribution of the longitudinal outcome. The main family of models available is generalized linear mixed models (GLMM) that includes any model from the exponential family. Additionally, it is possible to account for ordinal longitudinal outcomes through a proportional odds model, zero-inflated counts can be fitted with various zero-inflated models (i.e., zero-inflated Poisson, zero-inflated binomial, negative binomial, and betabinomial). Finally, zero-inflated continuous outcomes can be fitted with a two-part approach that consists in the decomposition of the outcome into two parts: the probability of non-zero value and the distribution of non-zero values conditional on a non-zero value. Various illustrative examples are described in the vignette of the package, available by running in \proglang{R} the command \code{vignette("INLAjoint")}.

The \code{formLong} argument corresponds to the formula for the mixed effects model. The structure of the formula is similar to the commonly used structure from the \proglang{R} package \pkg{lme4} \citep{Bates15}, which includes random effects as (NAME | ID), where NAME is the column that contains the variable to weight the random effects (1 corresponding to random intercept) and ID is the name of the grouping variable (e.g., individual id). In case of multiple longitudinal outcomes, a list of formulas can be provided (also in the case of a two-part mixed effects model as illustrated in \citet{Alvares22}). Note that the vector of random effects for a given longitudinal submodel have an unspecified correlation structure (i.e., full correlation between all random effects) but can be switched to independent random effects with the boolean argument \code{corRE} described below. Functions of time can be included in formulas, they first need to be set up as a univariate function with name fX, where X is a number starting with 1 and incrementing for each additional function of time required. Then the function can be used directly in the formula. See the second usage example for an illustration with splines. While it is possible to apply the function of time before calling the \code{joint} function, it is important to provide the function in the context of joint modeling where the linear predictor of the longitudinal models, including functions of time, can be shared in a survival submodel because the shared part needs to be computed at many time points internally to properly account for its evolution over time in the risk function.
%INLA internally uses the Cholesky for random effects. It is however possible to sample the standard deviations and correlations with \code{inla.iidkd.sample} (the argument \code{return.cov} allows to return samples of the covariance matrix instead).

The \code{dataLong} argument is the dataset that must contain the variables given in \code{formLong}. When multiple longitudinal outcomes are included, it is possible to supply only one dataset assuming all the variables from all formulas are present in this dataset and it is also possible to provide a list of datasets specific for each outcome (in the same order as the list of formulas).

Argument \code{id} is the name of the variable for repeated measurements (e.g., individuals). This is an unique string that should be the same for all outcomes. The \code{timeVar} argument gives the name of the time variable for longitudinal models.

The \code{family} argument is a character string (or a vector) giving the name of families for the longitudinal outcomes. The list of the available families can be displayed in \proglang{R} after loading the \pkg{INLA} library with the command inla.models()\$likelihoods. Some likelihoods have not been incorporated in \pkg{INLAjoint}, but all available likelihoods can be added on request.

The \code{link} argument is a character string (or a vector) giving the link function associated to the families for the longitudinal outcomes. The various links available for a given family can be displayed in \proglang{R} with the following command after loading the INLA library: \code{inla.doc("FAMILY NAME")}. The link should be a vector of the same size as the \code{family} parameter and should be set to \code{"default"} for default (i.e., identity for gaussian, log for poisson, logit for binomial, ...).

The \code{corLong} argument is a boolean (default is FALSE) that is only used when multiple longitudinal outcomes are included. When set to TRUE, the correlation structure between random effects accross longitudinal markers is unspecified, while random effects accross longitudinal markers are assumed independent when this is set to FALSE (i.e., block-diagonal correlation structure of the random effects). Similarly, the random effects for a given mixed effects model are assumed correlated by default but can be made independent. The \code{corRE} argument is a list of the size of number of groups of random effects (i.e., equal to 1 if there is only one longitudinal marker or if \code{corLong} is TRUE and equal to the number of markers otherwise), each element is a boolean indicating if the random effects of the group must be correlated or independent (i.e., diagonal variance-covariance).

The prior distributions of fixed and random effects can be specified within the \code{control} parameter, which is defined as a list with following entries relevant for longitudinal submodels:
\begin{itemize}
\item \code{priorFixed}: list with mean and standard deviations for the Gaussian prior distribution of the fixed effects. Default is \code{list(mean=0, prec=0.01, mean.intercept=0, prec.intercept=0.01)}, where \code{mean} and \code{prec} are the mean and precision (i.e., inverse of the variance) of the fixed effects, respectively and \code{mean.intercept} and \code{prec.intercept} are the corresponding parameters for the fixed intercept.
\item \code{priorRandom}: list with prior distribution for the multivariate random effects (inverse-Wishart). Default is \code{list(r=10, R=1)}, see \code{inla.doc("iidkd")} for more details.
\item{fixRE}: This argument allows to fix the variance and covariance of random effects when prior knowledge is available, which reduces the compuational burden. It is a list of the size of the number of groups of random effects, each element is a boolean indicating if the random effects of the group must be fixed or estimated.
\item{initVC}: This argument allows to set initial values for the variance and covariance of random effects. It is a list of the size of the number of groups of correlated random effects (i.e., size 1 when only one longitudinal marker is included or when \code{corLong} is set to TRUE), first values are variances and then covariances (same order as displayed in summary, for example for 3 correlated random effects the vector should be: Var1, Var2, Var3, Cov12, Cov13, Cov23, where ``Var'' are variances and ``Cov'' are covariances terms). All the elements of the covariance matrix must be fixed but in case of multiple groups of correlated random effects, it is possible to fix only some groups, then elements in the list that are not fixed must be an empty string.
\item{initSD}: This argument is the same as initVC but allows the user to fix standard deviations and correlations instead.
\end{itemize}

\subsection{Survival submodels}
The survival models are defined by proportional hazards models with parametric or semi-parametric baseline hazard approximation (where the AFT model can be defined with Weibull baseline as illustrated in \citet{Alvares22}).

The argument \code{formSurv} corresponds to the formula or the list of formulas for the time-to-event outcome(s), with the response given as an \code{inla.surv()} object which has the same form as \code{Surv()} from the \pkg{survival} package \citep{Therneau23}. The outcome can handle right, left and interval censoring as well as left truncation (for parametric baseline hazards). It is possible to fit a mixture cure model, where the predictors for the cure rate are also specified in the \code{inla.surv()} object for the outcome. Note that these predictors for the cure fraction are limited to fixed effects. See \code{?inla.surv} in \proglang{R} after loading the \pkg{INLA} library for more details. As for longitudinal submodels, it is possible to include random effects within parenthesis, which allows to fit frailty models for example.

The \code{dataSurv} argument should be a dataset that must contain the variables given in \code{formSurv}. When including multiple survival submodels, it is possible to supply only one dataset assuming all the variables from all formulas are present in this dataset and it is also possible to provide a list of datasets specific for each outcome (in the same order as the list of formulas). When fitting a joint model with a longitudinal component, if \code{dataSurv} is not provided, the longitudinal dataset is used to get the covariates values included in the time-to-event formula.
The \code{basRisk} argument corresponds to the baseline risk of event. It can be defined as parametric with either \code{"exponentialsurv"} for exponential baseline or \code{"weibullsurv"} for Weibull baseline (note that there are two formulations of the Weibull distribution, see \code{inla.doc("weibull")} for more details, the default is \code{variant = 0} but one can switch to the alternative formulation by including \code{variant = 1} in the list of control arguments). Alternatively, there are two options to avoid parametric assumptions on the shape of the baseline risk: \code{"rw1"} for a random walk of order one that corresponds to a smooth spline function based on first order differences. The second option \code{"rw2"} specifies a random walk order two that corresponds to a smooth spline function based on second order differences. This second option provides a smoother spline compared to order one since the smoothing is then done on the second order. 
The argument \code{NbasRisk} is the number of intervals for the decomposition of the follow-up, only one value should be provided and the same number of intervals is used for each survival submodel. Note that this decomposition is also used to account for time-dependent associations when fitting a longitudinal-survival joint model.
The \code{cutpoints} argument is a vector of values to manually define cutpoints if not using the default equidistant cutpoints (if not NULL, this replaces the \code{NbasRisk} parameter).
The \code{assocSurv} argument is a boolean that indicates if a frailty term (i.e., random effect) from a survival model should be shared and scaled into another survival model (i.e., joint shared frailty model, see \cite{rondeau07}). The order is important, the first model in the list of survival formulas (\code{formSurv}) should include a random effect and it can be shared in the next formulas. Multiple survival models with random effects can be accomodated and a random effect can be shared in multiple survival models (i.e., this argument should be a vector of booleans if one random effect is shared in multiple survival submodels and a list of vectors if multiple survival models with random effects share their random effects in multiple survival models).

\subsection{Joint modeling}
Some additional parameters are related to the joint modeling of longitudinal and survival outcomes, described here.
The \code{assoc} argument is a character string that specifies the association between the longitudinal and survival components. The available options are:
\begin{itemize}
\item \code{"CV"} for sharing the current value of the linear predictor (sometimes referred to as ``current level''), which corresponds to the value of the longitudinal linear predictor at a given time $t$ assumed to influence the hazard of event at the same time $t$.
\item \code{"CS"} for the current slope, corresponding to the derivative of the linear predictor at time $t$ with regard to time (i.e., the rate of change of the longitudinal marker) assumed to have an effect on the hazard of event at the same time $t$.
\item \code{"CV\_CS"} for the current value and the current slope
\item \code{"SRE"} for shared random effects (i.e., sharing the individual deviation from the mean at time t as defined by the random effects).
\item \code{"SRE\_ind"} for shared random effect independently (each random effect's individual deviation is associated to a scaling parameter in the survival submodel).
\item \code{""} (empty string) for no association
\end{itemize}
When there are either multiple longitudinal submodels or multiple survival submodels, this should be a vector. In case of both multiple markers and events, it should be a list with one element per longitudinal submodel and each element is a vector containing the association of this marker with each survival submodel.

Some additional \code{control} arguments are introduced here, the \code{control} argument is a list where the following entries can be placed:
\begin{itemize}
\item \code{priorAssoc}: a list with mean and standard deviations for the Gaussian prior distribution for the association parameters (does not apply to \code{"SRE\_ind"} association and shared random effect from survival models (frailty), see next item for those two). Default is list(mean=0, prec=0.01).
\item \code{priorSRE\_ind}: a list with mean and standard deviations for the Gaussian prior distribution on the association of independent random effects (\code{"SRE\_ind"} and survival frailty random effects shared). Default is list(mean=0, prec=1). The reason why these association parameters have their own prior is based on performances in simulations, these parameters are associated to constant variables while other association parameters described in the previous item are associated to variables that can be time varying, they are handled differently internally.
\item \code{assocInit}: Initial value for all the association parameters (default is 0.1).
\end{itemize}

An important point for joint models is the ability to handle any function of time in the longitudinal part and subsequently in the shared parts, a set of univariate functions of time is created and used in the longitudinal formulas. The reason why this is needed is because when sharing the linear predictor (CV), the deviation from the mean (SRE) or the derivative of the linear predictor (CS) that contains functions of time, we need to evaluate these function of time at various time points to integrate out the risk function in the likelihood. Moreover, when sharing the derivative of the linear predictor, \pkg{INLAjoint} uses the \proglang{R} package \pkg{numderiv} \citep{Gilbert19} to numerically approximate the derivative of the linear predictor at any time point.

\subsection{Additional informations}

Missing values in the outcomes for both longitudinal and survival submodels are handled properly by \pkg{INLAjoint}. The package uses the available information to replace these missing values with the posterior mean automatically in the joint modeling process.
 
Some additional \code{control} arguments that affect all models (i.e., longitudinal, survival and joint models) are described here.
\begin{itemize}
\item \code{int.strategy}: a character string giving the strategy for the numerical integration used to approximate the marginal posterior distributions of the latent field. Available options are \code{"ccd"} (default), \code{"grid"} or \code{"eb"}. The first two options are fully Bayesian and accounts for uncertainty over hyperparameters by using the mode and the curvature at the mode with points chosen with central composite design \citep{Box07} or a grid while the last one, the empirical Bayes strategy, only uses the mode from the first step of INLA's algorithm, it speeds up and simplifies computations. It can be pictured as a tradeoff between Bayesian and frequentist estimation strategies. Simulation studies showed that the speed up in computation time using empirical Bayes strategy does not degrade the frequentist properties of the fitted models \citep{Rustand23}.
\item \code{Ntrials}: Number of trials for binomial and Betabinomial distributions, default is NULL.
\item \code{cpo}: boolean with default as FALSE, when set to TRUE the Conditional Predictive Ordinate \citep{Pettit90} of the model is computed and returned in the model summary.
\item \code{cfg}: boolean with default as FALSE, set to TRUE to save configurations (which allows to sample from the full posterior using inla.posterior.sample() function. It is although possible to sample from the posterior when this option is kept as FALSE as described in the following.
\item \code{safemode}: boolean with default as TRUE (activated). Use the INLA safe mode (automatically reruns in case of negative eigenvalue(s) in the Hessian, reruns with adjusted starting values in case of crash). The message \code{"*** inla.core.safe:"} appears when the safe mode is triggered, it improves the inference of the hyperparameters when unstability is detected. To remove this safe mode, switch the boolean to FALSE (it can save some computation time but may return slightly less precise estimates for some hyperparameters when the model is unstable due to a lack of data observations, misspecification or identifiability issues).
\item \code{rerun}: boolean with default as FALSE. Force the model to rerun using posterior distributions of the first run as starting values of the new run. Most of the time it will not change the results but it can improve numerical stability for unstable models.
\item \code{tolerance}: accuracy in the inner optimization (default is 0.005).
\item \code{h}: step-size for the hyperparameters (default is 0.005).
\item \code{verbose}: boolean with default as FALSE, prints details of the INLA algorithm when set to TRUE. 
\end{itemize}

Of note, \pkg{INLAjoint} uses \code{OPENMP} \citep{Dagum98} for parallel computations. Since parallelisation of computations is done on a non-predefined order, running the same model twice may lead to slightly different results following aggregation of results. However, it is possible to run sequentially by setting the option \code{inla.setOption(num.threads="1:1")} before calling the \code{joint} function. Moreover, INLA internally uses some optimisation that can slightly differ from one run to another and it is possible to fix this optimization process by setting \code{internal.opt=FALSE} in \code{control} options of the call of the \code{joint} function, although the results will be almost identical when keeping these options as default unless the model is very unstable (identifiability issues or very few data observations).

\subsection{S3 methods}
The primary objective of the \pkg{INLAjoint} \proglang{R} package is to provide a flexible and versatile framework for constructing complex regression models, specifically designed to address the challenges posed by multivariate outcomes. Unlike alternative software which applies to a limited range of models and offer poor scaling for complex models (many outcomes, parameters and/or data), \pkg{INLAjoint} is designed to be flexible, allowing users to integrate diverse regression models and association structures to address nuanced research questions that involve both longitudinal and survival outcomes. In this context, the \code{summary}, \code{plot} and \code{predict} functions are tailored to the analysis of longitudinal and survival data, enhancing the interpretability and usability of the fitted joint models. This is an important feature as the INLA methodology implemented in the R package \pkg{INLA} is not designed for these models specifically and require significant post-processing to calculate the necessary and standard joint modeling quantities. 

First, the \code{summary} function is returning results grouped by outcomes and allows to choose the metrics for residual error terms of Gaussian and lognormal models between variance and standard deviation while INLA uses precision for computational convenience. Similarly, the random effects are parametrized through the cholesky matrix but returned as either variance-covariance or standard deviation and correlations. The boolean argument \code{sdcor} in the \code{summary} function allows to switch between the two options for displaying the output. For survival submodels, the boolean \code{hr} allows to convert fixed effects estimates into hazard ratios. These features allow the user to choose the desired scale of the output while most software impose a scale and require cumbersome post-computation and sampling to switch the scale and get uncertainty quantification on this scale. Additionally, the marginal log-likelihood and goodness-of-fit metrics are provided in the output of \code{summary}, including the DIC (deviance information criterion, \cite{spiegelhalter02}) and WAIC (widely applicable Bayesian information criterion, \cite{watanabe13}). 

The plot function, when applied to an object fitted with \pkg{INLAjoint} returns the marginal posterior distributions of all the model parameters as well as baseline hazard parameters and/or baseline hazard curves. It is also possible to switch from variance and covariance of residual error terms and random effects to standard deviations and correlations with the argument \code{sdcor}. The boolean \code{priors} allows to add the prior distribution on each posterior distribution plot as illlustrated in the second usage example (Section \ref{UseEx}), which is very useful to identify parameters that may rely mostly on the prior (i.e., the observed data is not informative) and to do prior sensitivity analysis to assess the impact of priors on the posteriors. It is not always easy to understand the implication of priors in Bayesian inference and this tool allows for visualisation of their distribution and impact on the results. Evaluating the impact of priors on the results of a Bayesian model by looking at the posterior obtained from different priors can avoid wrong conclusion due to misleading priors and this tool facilitates this process.

The \code{predict} function has multiple purposes, it uses a simple syntax that requires a fitted model and new data to compute the model predictions for all the outcomes included in a model, based on the new data. It can be used for imputation to obtain the expected value of the model at any time point, for forecasting to get the predicted trajectory of an outcome and for inference as it allows to compute average trajectories for given covariates. The \code{horizon} argument allows to choose the horizon of the prediction. With default options, the predict function returns the value of the linear predictor for each individual in the new data provided at a number of time points that can be defined with argument \code{NtimePoints} (default 50). It is however possible to choose the time points manually with the argument \code{timePoints}. Uncertainty is quantified through sampling, where the number of samples for fixed effects and hyperparameters is defined by the argument \code{Nsample} (default 300). For each sample, random effects realisations are sampled conditional on the observed longitudinal data in the new dataset provided, the number of random effects nested samples can be chosen with the argument \code{NsampleRE} (default 50). Since we sample random effects conditional on the observed longitudinal data, the number of required samples to properly measure uncertainty is reduced compared to the usual approach that samples from the marginal distribution of random effects. The default output returns summary statistics over the linear predictor of each outcome but it is possible to get the sampled curves instead by switching the boolean argument \code{return.samples} to TRUE.

For a longitudinal outcome, the argument \code{inv.link} allows to apply the inverse link function in order to get predictions on an interpretable scale. For example for a binomial model for binary data with a logit link, when \code{inv.link} is set to TRUE in the call of \code{predict}, the returned predictions give the probability of 1 versus 0 instead of the predicted linear predictor, with uncertainty quantification on this scale. This feature allows to directly interpret and display the results and the observed data as illustrated in our second usage example (Section \ref{UseEx}). For survival submodels, with default options the prediction assume the individual did not experience the survival event(s) before the last longitudinal observation provided in the new dataset. However, it is possible to manually define the time point at which survival prediction should start with the argument \code{Csurv}. The boolean argument \code{survival} alllows to compute summary statistics over survival curves in addition to the risk curves. When fitting a competing risks model, the boolean argument \code{CIF} allows to compute the cumulative incidence function instead of survival, which is more interesting to interpret as survival curves represents the probability of having an event in a hypothetical world where it is not possible to have any of the competing events while the cumulative incidence functions give the probability of observing an event accounting for the fact that subjects may have one of the competing events before. See the second usage example for an illustration (Section \ref{UseEx}).

While it is possible to use the predict function to predict the longitudinal and survival models trajectories over new individuals or forecast for some known individuals, it also allows to predict the average trajectory for any outcome conditional on covariates. In this case, the new data should only consist of one line that gives the value of covariates and the longitudinal outcomes values should be set to \code{NA}. The predict function will automatically assume the average trajectory of each outcome conditional on the given covariates, and will quantify uncertainty through sampling where the marginal distribution of random effects is used (instead of the distribution conditional on observed longitudinal measurements when provided) which allows to display the average trajectories for inference purposes. This is very useful for complex models as it is common to use models that allow for complex trajectories (e.g., fixed and random effects on splines or complex interactions), where the value of parameters is difficult to directly interpret. With this feature, one can compute the average trajectory conditional on treatment and compare the longitudinal and survival profiles of each treatment line visually for example. This feature is illustrated in the second usage example (Section \ref{UseEx}). Note that predictions with \pkg{INLAjoint} are fully Bayesian, meaning that all the parameters are sampled (i.e., including hyperparameters) to assess uncertainty.

It is possible to sample marginalized fixed effects realizations of any model (i.e., hyperparameters integrated out) with the function \code{inla.rjmarginal(N, MODEL)}, where \code{N} is the number of samples and \code{MODEL} is a fitted model with \pkg{INLAjoint}. To sample hyperparameters, the function \code{inla.hyperpar.sample(N, MODEL)} can be used similarly.

\section[Usage Examples]{Usage Examples}
\label{UseEx}
We illustrate the usage of \pkg{INLAjoint} through two examples of multivariate joint modeling. The first one shows how to define a joint model with 2 longitudinal outcomes and a survival outcome to introduce basic concepts of \pkg{INLAjoint}'s usage and compare with alternative software. The second example illustrates advanced features in a model that includes 7 longitudinal outcomes and two competing risks of events, this complex joint model cannot be fitted with alternative software in the literature at the moment. All the computations are done over 12 Intel Xeon Gold 6248 2.50GHz CPUs. We used \proglang{R} version 4.3.2, \pkg{INLAjoint} version 24.3.25, \pkg{INLA} version 24.03.09, \pkg{JMbayes2} version 0.4-5 and \pkg{rstanarm} version 2.32.1.

\subsection{pbc2 dataset}
We use the pbc2 dataset as provided in the \proglang{R} package \pkg{JMbayes2}. This dataset containts longitudinal information of 312 randomised patients with primary biliary cirrhosis disease, followed at the Mayo Clinic between 1974 and 1988 \citep{murtaugh94}. We considered 7 longitudinal markers: SGOT (aspartate aminotransferase in U/ml., lognormal), platelets per cubic ml/1000 (Poisson) serum bilirubin (in mg/dl, lognormal), albumin in g/dl (Gaussian), ascites (no/yes, binomial), spiders (no/yes, binomial), prothrombine time in seconds (Gaussian). Some patients died during follow-up (140) and 29 received liver transplantation, which we will consider as a competing terminal event. The maximum follow-up time is 14.3 years with a number of individual repeated measurements ranging between 1 and 16 with a median of 5. This dataset has been widely used to illustrate joint modeling approaches for multivariate longitudinal and survival data \citep{Rustand23, hughes23, devaux22, murray22, philipson20} but a joint model for 9 outcomes (7 longitudinal and 2 survival) as we propose in the second example, has not been proposed or implemented yet.

\subsection{Example 1: Joint model for two longitudinal outcomes and a terminal event}
\label{Ex1}
In this first example, we model two longitudinal outcomes with a lognormal mixed effects model and a Poisson mixed effects model, both including fixed effects for the intercept, slope and drug as well as correlated random intercepts. Let $Y_{i1}(t)$ denote the value of SGOT (i.e., aspartate aminotransferase) for individual $i$ at time $t$ and $Y_{i2}(t)$ the corresponding value for the platelet counts. Assuming $k=1,2$ for SGOT and platelet, respectively, the fixed intercept is denoted by $\beta_{k0}$, the fixed slope by $\beta_{k1}$, the effect of drug at baseline by $\beta_{k2}$ and the effect of drug over time by $\beta_{k3}$. Each longitudinal submodel includes a Gaussian random intercept with variance $\sigma^2_k$ and their covariance is denoted $\sigma_{12}$. The residual error term for the lognormal observations is denoted by $\varepsilon_{i1}(t)$, which is assumed Gaussian with mean 0 and standard deviation $\sigma_\varepsilon$. The risk of the composite event (i.e., death or transplantation) for individual $i$ at time $t$ is denoted by $\lambda_i(t)$ and defined by a baseline risk $\lambda_0(t)$ and a linear predictor including a fixed effect of drug $\gamma_1$ and the effect of the shared linear predictor from the two longitudinal submodels $\varphi_1$ and $\varphi_2$. The purpose of this example is to compare \pkg{INLAjoint}'s fit with the two available alternatives in \proglang{R} able to fit this model, \pkg{JMbayes2} and \pkg{rstanarm}, in order to highlight \pkg{INLAjoint}'s computational speed and accuracy. We keep the default priors for all packages despite their differences as the intention is to compare packages as they are meant to be used. Simulation studies evaluated the differences in the case of matching priors between \pkg{INLAjoint} and \pkg{rstanarm} \citep{Rustand23}. Note that \pkg{JMbayes2} and \pkg{rstanarm} use LME from univariate models (i.e., mixed effects and proportional hazards models) to define initial values and \pkg{JMbayes2} also uses those univariate models to define data-driven prior distributions for the joint model fit. The three packages assume Gaussian priors for the fixed effects, where \pkg{rstanarm} assume a variance of 2.5 while \pkg{INLAjoint} assume a variance of 100. For the random effects covariance matrix, \pkg{rstanarm} uses the LKJ prior while \pkg{JMbayes2} and \pkg{INLAjoint} use inverse-Wishart. More details about the priors and initial values used with \pkg{INLAjoint} can be displayed from a fitted model with the function \code{inla.priors.used(MODEL)}, where \code{MODEL} is the name of the object that contains the fitted model. Moreover, the 3 methods use different approximation of the baseline hazard, \pkg{rstanarm} uses cubic B-splines for the log baseline hazard with 2 internal knots while \pkg{JMbayes2} uses quadratic B-splines with 9 internal knots. With \pkg{INLAjoint}, we illustrate the data augmentation of a Weibull baseline hazard where the follow-up is decomposed into 15 intervals with equidistant nodes (betwen 0 and maximum observed time). Therefore, the comparison will not focus on the baseline hazard estimation as it is handled differently for each package. The default number of MCMC chains and iterations is used with \pkg{JMbayes2} (i.e., 6500 iterations and 3 chains for this model) but we increased the default number of iterations with \pkg{rstanarm} to 6000 with 4 chains instead of the default 2000 iterations with 4 chains in order to reach proper convergence as results with less iterations were unstable. The fitted model is defined as follows:
\[
\left\{
\setlength\arraycolsep{0pt}
\begin{array}{ c @{{}{}} l  @{{}{}} r }
\log(Y_{i1}(t)) &= \eta_{i1}(t) + \varepsilon_{i1}(t) & \hspace{-5cm} \textbf{\textit{(aspartate aminotransferase (SGOT) - lognormal)}}\\
&=  \beta_{10} + b_{i1} +\beta_{11}t+\beta_{12}drug_{i} + \beta_{13}tdrug_i + \varepsilon_{i1}(t) \\ \ \\

\log(E[Y_{i2}(t)])&=\eta_{i2}(t) & \textbf{\textit{(platelet - Poisson)}}\\
&=  \beta_{20} + b_{i2} +\beta_{21}t+\beta_{22}drug_{i} + \beta_{23}tdrug_i\\ \ \\

\lambda_{i}(t)&=\lambda_{0}(t)\ \textrm{exp}\left(\gamma_1 drug_i + \eta_{i1}(t)\varphi_{1} + \eta_{i2}(t)\varphi_{2} \right) & \textbf{\textit{(event risk)}}\\ \ \\
\end{array}
\right.
\]
We start by loading the packages and the longitudinal (\code{pbc2}) and survival (\code{pbc2.id}) data, extracting the columns of interest for this first model.

\begin{CodeInput}
R> library(INLA)
R> library(INLAjoint)
R> library(JMbayes2)
R> library(rstanarm)
R> data("pbc2.id")
R> data("pbc2")
R> pbc2_1 <- pbc2[, c("id", "drug", "year", "SGOT", "platelets")]
\end{CodeInput}
We can then fit the joint model with \pkg{INLAjoint}:
\begin{CodeChunk}
\begin{CodeInput}
R> IJ <- joint(formSurv = list(inla.surv(years, status2) ~ drug),
+              formLong = list(SGOT ~ year * drug + (1|id),
+                              platelets ~ year * drug + (1|id)),
+              dataLong = pbc2_1, dataSurv = pbc2.id, id = "id",
+              corLong=TRUE, timeVar = "year", basRisk="weibullsurv",
+              family = c("lognormal", "poisson"), assoc = c("CV","CV"))
R> summary(IJ, sdcor=TRUE)
\end{CodeInput}
\begin{CodeOutput}
Longitudinal outcome (L1, lognormal)
                        mean     sd 0.025quant 0.5quant 0.975quant
Intercept_L1          4.7971 0.0388     4.7210   4.7971     4.8733
year_L1              -0.0051 0.0038    -0.0126  -0.0051     0.0024
drugDpenicil_L1      -0.1545 0.0547    -0.2618  -0.1545    -0.0472
year:drugDpenicil_L1 -0.0014 0.0053    -0.0118  -0.0014     0.0091
Res. err. (sd)        0.3070 0.0053     0.2967   0.3069     0.3175

Longitudinal outcome (L2, poisson)
                        mean     sd 0.025quant 0.5quant 0.975quant
Intercept_L2          5.5102 0.0319     5.4475   5.5102     5.5728
year_L2              -0.0478 0.0009    -0.0495  -0.0478    -0.0461
drugDpenicil_L2      -0.1014 0.0449    -0.1894  -0.1014    -0.0133
year:drugDpenicil_L2  0.0138 0.0012     0.0114   0.0138     0.0162

Random effects standard deviation / correlation
                             mean     sd 0.025quant 0.5quant 0.975quant
Intercept_L1               0.4473 0.0198     0.4105   0.4466     0.4870
Intercept_L2               0.3947 0.0158     0.3648   0.3943     0.4273
Intercept_L1:Intercept_L2 -0.1382 0.0605    -0.2545  -0.1380    -0.0158

Survival outcome
                     mean     sd 0.025quant 0.5quant 0.975quant
Weibull (shape)_S1 1.0639 0.0730     0.9532   1.0541     1.2347
Weibull (scale)_S1 0.0420 0.0242     0.0137   0.0376     0.0950
drugDpenicil_S1    0.1116 0.1715    -0.2246   0.1116     0.4480

Association longitudinal - survival
            mean     sd 0.025quant 0.5quant 0.975quant
CV_L1_S1  1.3724 0.2184     0.9510   1.3695     1.8110
CV_L2_S1 -1.1338 0.2072    -1.5435  -1.1331    -0.7277

log marginal-likelihood (integration)    log marginal-likelihood (Gaussian)
                            -45022.14                             -45022.14

Deviance Information Criterion:  8725.475
Widely applicable Bayesian information criterion:  15606.05
Computation time: 15.24 seconds
\end{CodeOutput}
\end{CodeChunk}
The results are displayed with the \code{summary} function, starting with parameters related to the first longitudinal outcome (SGOT), including fixed effects and the standard deviation of the residual error. Then the parameters corresponding to the second longitudinal outcome are displayed and below is the variance-covariance of random effects. Adding the argument \code{sdcor=TRUE} to the call of the summary function returns standard deviations of the residual error and the random effects as well as correlation between random effects instead of the default variance and covariance as illustrated in the call. Parameters related to the survival outcome are then displayed, with the shape and scale of the Weibull distribution for the baseline hazard and the fixed effect of drug. Finally, the association parameters for the current value of the two shared linear predictors to scale their effect on the risk of event are displayed. Parameters related to the risk of event can be transformed to hazard ratios by adding the argument \code{hr=TRUE} (default is FALSE) to the call of the summary function, which properly handles uncertainty. The names of outcomes are conveniently named with the letter ``L" for longitudinal outcomes and ``S" for survival outcomes followed by their position in the list of formulas, allowing to easily identify the involved components for association parameters and when plotting and doing predictions. The marginal likelihood of the model and some goodness of fit criteria are provided at the end of the summary function (i.e., DIC, WAIC).

\begin{figure}[!ht]
\centering
\includegraphics[scale=1]{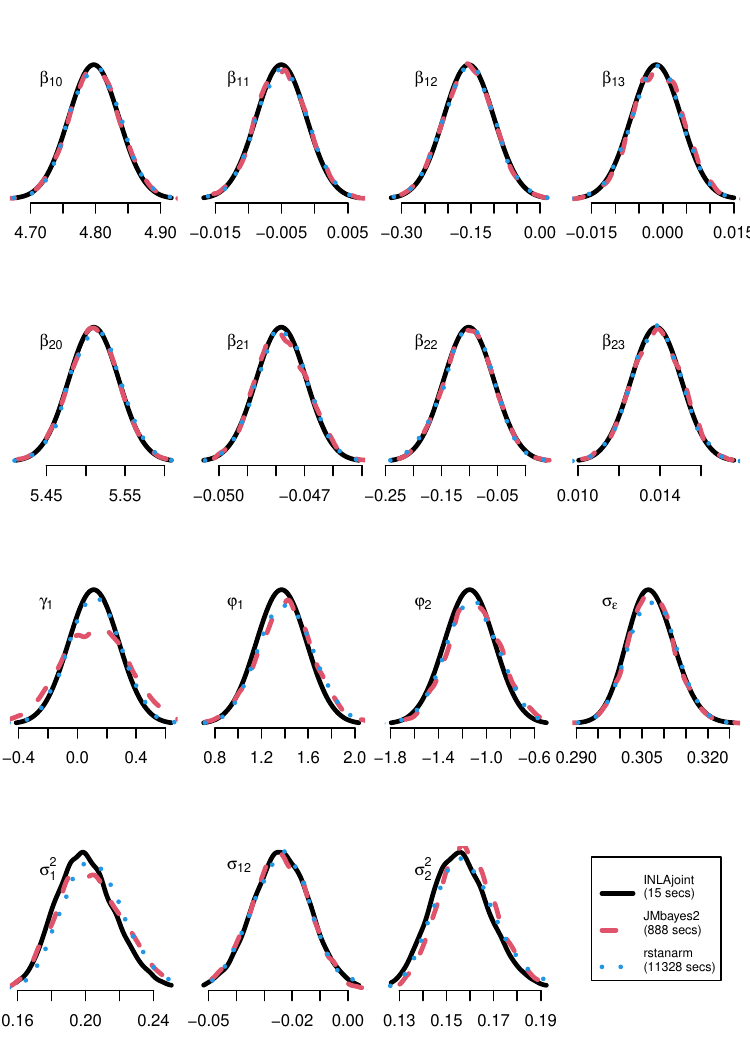}
\caption{Marginal posterior distributions for the model described in Section \ref{Ex1} with \pkg{INLAjoint}, \pkg{JMbayes2} and \pkg{rstanarm}. Computation time is provided in the legend.}
\label{DensityPlot}
\end{figure}

We then fit a similar model with \pkg{JMbayes2} and \pkg{rstanarm}:
\begin{CodeInput}
R> a_JB2 <- Sys.time()
R> fm1 <- lme(fixed = log(SGOT) ~ year * drug, random = ~ 1 | id, data = pbc2_1)
R> fm2 <- mixed_model(platelets ~ year * drug, data = pbc2_1,
+                     random = ~ 1 | id, family = poisson())
R> Mixed <- list(fm1, fm2)
R> fCox1 <- survreg(Surv(years, status2) ~ drug, data = pbc2.id)
R> JB2 <- jm(fCox1, Mixed, time_var = "year", control=list(cores=12))
R> b_JB2 <- Sys.time()
R> JB2_CT <- round(difftime(b_JB2, a_JB2, units="secs"))
R> JB2_mar <- apply(do.call(cbind, sapply(JB2$mcmc[c(8,9,3,4,10,7)],
+                                         function(x) do.call(rbind, x))),
+                   2, density)
R> JB2_mar <- append(JB2_mar, JB2_CT)

R> RS <- stan_jm(
+   formulaLong = list(SGOT ~ year * drug + (1|id),
+                       platelets ~ year * drug + (1|id)), dataLong = pbc2_1,
+   formulaEvent = survival::Surv(years, status2) ~ drug, dataEvent = pbc2.id,
+   family = list(gaussian(link=log), poisson), chains = 4, iter = 6000,
+   time_var = "year", seed = 12345, cores=12)
R> RS_mar <- apply(as.matrix(RS)[, c(1, 3:5, 2, 6:8, 10:12, 637, 644:646)],
+                  2, density)
R> RS_mar <- append(RS_mar, max(RS$runtime)*60)
\end{CodeInput}
The computation time is manually recorded with \pkg{JMbayes2} to be able to fairly compare as it only stores computation for the last step of the joint model fit by default (while the fitting procedure involves fitting each univariate model before fitting the joint model). We also extract the posterior marginal distributions from each model fit to compare the distribution of parameters with density plots. We take advantage of the plot function for \pkg{INLAjoint} as it automatically computes posterior marginals on the desired scale (e.g., standard deviation instead of precision for residual error and covariances for random effects), which we can conveniently extract to match the scale of the two other packages's marginals:

%we take advantage of the plot function that computes marginal distributions on the desired scale with INLAjoint as they are internally parametrized differently (precision instead of variance of residual error and cholesky matrix of random effects instead of variance-covariance)

\begin{CodeChunk}
\begin{CodeInput}
R> IJp <- plot(IJ)
R> IJ_mar <- c(split(IJp$Outcomes$L1$data, ~Effect)[2:5],
+              split(IJp$Outcomes$L2$data, ~Effect)[6:9],
+              split(IJp$Outcomes$S1$data, ~Effect)[1],
+              split(IJp$Associations$data, ~Effect),
+              split(plot(IJ, sdcor=T)$Outcomes$L1$data, ~Effect)[10],
+              split(IJp$Covariances$L1$data, ~Effect)[-3])

R> legend_text <- expression(beta[10], beta[11], beta[12], beta[13],
+                            beta[20], beta[21], beta[22], beta[23],
+                            gamma[1], varphi[1], varphi[2],
+                            sigma[epsilon], sigma[1]^2, sigma[12], sigma[2]^2)

R> Fplot <- function(x){
+   plot(IJ_mar[[x]]$x, IJ_mar[[x]]$y, type = "l", xlab = "", ylab = "", 
+        main = "", ann=FALSE, bty="n", yaxt='n', lwd=3)
+   lines(JB2_mar[[x]]$x, JB2_mar[[x]]$y, col=2, lty=2, lwd=3)
+   lines(RS_mar[[x]]$x, RS_mar[[x]]$y, col=4, lty=3, lwd=3)
+   legend("topleft", legend_text[x], bty = "n")
+  }

R> pdf("DensityPlot.pdf", width=5, height=7)
R> layout.matrix <- t(matrix(c(c(1:4), c(5:8), c(9:12), c(13:16)), ncol=4))
R> layout(mat = layout.matrix)
R> par(mar = c(3,0.5,3,0.5), cex.lab=0.1)
R> sapply(1:15, Fplot)
R> plot.new()
R> legend("center", legend=c(paste0("INLAjoint\n(", round(IJ$cpu.used[4]),
+                                   " secs)\n"),
+                            paste0("JMbayes2\n(", JB2$CT, " secs)\n"),
+                            paste0("rstanarm\n(", max(RS$runtime)*60,
+                                   " secs)\n")),
+         lty=c(1,2,3), lwd=c(3,3,3), col=c(1,2,4), seg.len=3, cex=0.7)
R> dev.off()
\end{CodeInput}
\end{CodeChunk}
The resulting plot is presented in Figure \ref{DensityPlot}. We can see that \pkg{INLAjoint} gives the same results as the MCMC packages but within 15 seconds compared to 15 minutes and 190 minutes for \pkg{JMbayes2} and \pkg{rstanarm}, respectively. This illustrates how \pkg{INLAjoint} facilitates the application of joint models, as their use have been restrained by their computational burden, allowing for more complex models and big data applications at a reasonable computational cost. Additionally, simulation studies \citep{Rustand23} have shown how this difference in computation time scales in favor of INLA compared to MCMC for more complex models.

\subsection{Example 2: Joint model for 7 longitudinal outcomes and 2 competing risks of event}
\label{Ex2}
The second model illustrates many additional features available in \pkg{INLAjoint}. As far as we know, no other software can fit this model in the literature at the moment  since \pkg{rstanarm} is limited to 3 longitudinal markers and \pkg{JMbayes2} is limited to unconstrained covariance between random effects across markers. Moreover, the required computation time to reach convergence with MCMC for such complex models  becomes unreasonable as illustrated in \citep{Rustand23}. The model is defined as follows:

\[
\left\{
\setlength\arraycolsep{0pt}
\begin{array}{ c @{{}{}} l  @{{}{}} r }
\log(Y_{i1}(t)) &= \eta_{i1}(t) + \varepsilon_{i1}(t) \hspace{4cm} \textbf{\textit{(serum Bilirubin - lognormal)}}\\
&=(\beta_{10} + b_{i10})+(\beta_{11}+b_{i11})\textrm{NS}_1(t)+(\beta_{12}+b_{i12})\textrm{NS}_2(t) + \varepsilon_{i1}(t) \\ \ \\

\log(E[Y_{i2}(t)]) &= \eta_{i2}(t)  \hspace{7.5cm}  \textbf{\textit{(platelet - Poisson)}}\\
&= (\beta_{20} + b_{i20})+(\beta_{21}+b_{i21})t+\beta_{22}drug_i + \beta_{23}sex_i + \beta_{24}tdrug_i\\
& \ \ \ + \beta_{25}tsex_i + \beta_{26}drug_isex_i + \beta_{27}tdrug_isex_i \\ \ \\

Y_{i3}(t)&=\eta_{i3}(t) + \varepsilon_{i3}(t) \hspace{0.5cm}  \textbf{\textit{(aspartate aminotransferase (SGOT) - lognormal)}}\\
&= (\beta_{30} + b_{i30})+\beta_{31}t + \varepsilon_{i3}(t)\\ \ \\

Y_{i4}(t)&=\eta_{i4}(t) + \varepsilon_{i4}(t) \hspace{5.5cm}  \textbf{\textit{(albumin - Gaussian)}}\\
&= (\beta_{40} + b_{i40})+\beta_{41}t + \varepsilon_{i4}(t)\\ \ \\

\textrm{logit}(E[Y_{i5}(t)])&=\eta_{i5}(t) \hspace{7cm}  \textbf{\textit{(ascites - Binomial)}}\\
&= (\beta_{50} +b_{i50})+\beta_{51}t \\ \ \\

\textrm{logit}(E[Y_{i5}(t)])&=\eta_{i5}(t) \hspace{7cm}  \textbf{\textit{(spiders - Binomial)}}\\
&= (\beta_{60} +b_{i60})+\beta_{61}t \\ \ \\

Y_{i7}(t)&=\eta_{i7}(t) \hspace{6cm}  \textbf{\textit{(prothrombin - Gaussian)}}\\
&= (\beta_{70} +b_{i70})+\beta_{71}t \\ \ \\

\lambda_{i1}(t)&=\lambda_{01}(t)\ \textrm{exp}\left(\eta_{i1}(t)\varphi_{11} + \eta_{i2}(t)\varphi_{12} + b_{i50}\varphi_{13} + \frac{\partial \eta_{i7}(t)}{\partial t}\varphi_{14}\right) \hspace{0.5cm}   \textbf{\textit{(death risk)}}\\ \ \\

\lambda_{i2}(t)&=\lambda_{02}(t)\ \textrm{exp}\left(\eta_{i2}(t)\varphi_{21} + \eta_{i3}(t)\varphi_{22} + b_{i40}\varphi_{23} + \eta_{i6}(t)\varphi_{24} + \frac{\partial \eta_{i6}(t)}{\partial t}\varphi_{25}\right)\\ &\hspace{8.5cm}  \textbf{\textit{(transplantation risk)}}
\end{array}
\right.
\]
The first mixed effects model fits lognormal serum bilirubin levels with two natural cubic splines basis, each associated to random effects in addition to the random intercept. The second mixed effects model fits platelets counts with a Poisson distribution, it includes fixed and random intercept and slope and fixed effects for the triple interaction between time, drug and sex. The remaining mixed effects regression models corresponds to SGOT (lognormal), albumin (Gaussian), ascites (binomial), spiders (binomial) and prothrombin (Gaussian), all associated to fixed and random intercept as well as a fixed slope. The two remaining models are proportional hazards models for the risk of death and the risk of transplantation. They include components from the longitudinal mixed effects models. Shared linear predictors (i.e., current value) are denoted by $\eta_{ik}(t)$, where $k$ identifies the longitudinal model, shared random effects are denoted by $b_{ik}$ and the shared derivatives of the linear predictors (i.e., current slope) are denoted by $\frac{\partial \eta_{ik}(t)}{\partial t}$ and the corresponding scaling parameters are denoted by $\varphi$.

We start by loading the data (some additional columns are required compared to the previous model) and we scale the continuous outcomes for ease of interpretability and model integrity and we convert factors to integers to facilitate the plotting of the results (so we can display the probability of 1 vs. 0 and the observed data simultaneously in the illustration of the predict function).

\begin{CodeInput}
R> pbc2_2 <- pbc2[, c("id", "drug", "sex", "year", "ascites", "spiders",
+                     "serBilir", "albumin", "SGOT", "platelets", "prothrombin")]
R> pbc2_2[, c(7, 9, 11)] <- log(pbc2_2[, c(7, 9, 11)])
R> colnames(pbc2_2)[c(7,9,11)] <- c("log_serBilir", "log_SGOT", "log_prothrombin")
R> pbc2_2[, c(7:9, 11)] <- scale(pbc2_2[,c(7:9, 11)])
R> pbc2_2$drug <- as.integer(pbc2_2$drug)-1
R> pbc2_2$sex <- as.integer(pbc2_2$sex)-1
R> pbc2_2$ascites <- as.integer(pbc2_2$ascites)-1
R> pbc2_2$spiders <- as.integer(pbc2_2$spiders)-1
\end{CodeInput}

Then we create the two survival outcomes (we assume competing risks of death and transplantation instead of the composite outcome used in the first model). We also illustrate how to define user-created functions to have any transformation function of time that can be used for fixed and random effects in the formulas for the longitudinal models. These functions are used internally when integrating the hazard function to compute the survival function in the likelihood, when a time-dependent component is shared into a survival submodel.
\begin{CodeInput}
R> pbc2.id$death <- ifelse(pbc2.id$status=="dead", 1, 0)
R> pbc2.id$tsp <- ifelse(pbc2.id$status=="transplanted", 1, 0)
R> BSP <- ns(pbc2_2$year, knots=1)
R> f1 <- function(x) predict(BSP, x)[,1]
R> f2 <- function(x) predict(BSP, x)[,2]
\end{CodeInput}

We can then fit the model and print the results with the \code{summary} function. Note that the argument \code{corLong} is set to FALSE in order to have a block-diagonal covariance structure for random effects (i.e., independence between longitudinal markers), which reduces the number of parameters to fit and thus the computation time. It has been shown that this independence assumption can have minor impact on the other parameters estimates while reducing the model complexity \citep{Rustand23}, although the model can also be fitted with an unspecified covariance structure by switching this argument to TRUE. Moreover, the empirical Bayes strategy for the hyperparameters is used to reduce even further the computation time by setting the argument \code{int.strategy="eb"} in \code{control} options. It allows for faster computation without loss of accuracy and good frequentist properties as illustrated through simulation studies in \citet{Rustand23}. For the baseline hazard, we assume Bayesian smooth splines corresponding to random walk priors of order two and one for the baseline hazard of death and transplantation, respectively.

\begin{CodeChunk}
\begin{CodeInput}
R> IJ2 <- joint(formSurv = list(inla.surv(years, death) ~ 1,
+                               inla.surv(years, tsp) ~ 1),
+               formLong = list(serBilir ~  f1(year) + f2(year) +
+                                           (1 + f1(year) + f2(year) |id),
+                               platelets ~ year * drug * sex + (1 + year|id),
+                               SGOT ~ year + (1|id),
+                               albumin ~ year + (1|id),
+                               ascites ~ year + (1|id),
+                               spiders ~ year + (1|id),
+                               prothrombin ~ year + (1|id)),
+               dataLong = pbc2_2, dataSurv = pbc2.id, id = "id", corLong=FALSE,
+               timeVar = "year", basRisk=c("rw2","rw1"),
+               family = c("gaussian", "poisson", "gaussian", "gaussian",
+                          "binomial", "binomial", "gaussian"),
+               assoc = list(c("CV", ""), c("CV", "CV"), c("", "CV"), c("", "SRE"),
+                            c("SRE", ""), c("", "CV_CS"), c("CS", "")),
+               control=list(int.strategy="eb"))
R> summary(IJ2)
\end{CodeInput}
\begin{CodeOutput}
Longitudinal outcome (L1, gaussian)
                        mean     sd 0.025quant 0.5quant 0.975quant
Intercept_L1         -0.0784 0.0518    -0.1800  -0.0784     0.0232
f1year_L1             1.9748 0.1386     1.7030   1.9748     2.2465
f2year_L1             2.0532 0.1865     1.6877   2.0532     2.4186
Res. err. (variance)  0.0741 0.0030     0.0684   0.0740     0.0802

Random effects variance-covariance (L1)
                         mean     sd 0.025quant 0.5quant 0.975quant
Intercept_L1           0.8161 0.0784     0.6803   0.8102     0.9845
f1year_L1              3.9244 0.5821     2.9605   3.8636     5.2134
f2year_L1              3.5459 0.8087     2.1951   3.4625     5.3296
Intercept_L1:f1year_L1 0.5876 0.1700     0.2851   0.5747     0.9398
Intercept_L1:f2year_L1 0.5966 0.2836     0.0649   0.5777     1.1698
f1year_L1:f2year_L1    1.9361 0.6285     0.9121   1.8699     3.3822

Longitudinal outcome (L2, poisson)
                    mean     sd 0.025quant 0.5quant 0.975quant
Intercept_L2      5.4487 0.1000     5.2527   5.4487     5.6448
year_L2          -0.1166 0.0497    -0.2140  -0.1166    -0.0191
drug_L2          -0.0506 0.1310    -0.3074  -0.0506     0.2062
sex_L2            0.0814 0.1053    -0.1249   0.0814     0.2878
year:drug_L2     -0.0410 0.0652    -0.1687  -0.0410     0.0868
year:sex_L2       0.0595 0.0522    -0.0429   0.0595     0.1618
drug:sex_L2      -0.0222 0.1391    -0.2948  -0.0222     0.2504
year:drug:sex_L2  0.0028 0.0691    -0.1328   0.0028     0.1383

Random effects variance-covariance (L2)
                        mean     sd 0.025quant 0.5quant 0.975quant
Intercept_L2          0.1492 0.0118     0.1272   0.1487     0.1741
year_L2               0.0309 0.0042     0.0236   0.0306     0.0394
Intercept_L2:year_L2 -0.0041 0.0050    -0.0142  -0.0041     0.0061

Longitudinal outcome (L3, gaussian)
                        mean     sd 0.025quant 0.5quant 0.975quant
Intercept_L3          0.1127 0.0495     0.0156   0.1127     0.2098
year_L3              -0.0123 0.0048    -0.0217  -0.0123    -0.0029
Res. err. (variance)  0.3034 0.0107     0.2829   0.3032     0.3250

Random effects variance-covariance (L3)
               mean     sd 0.025quant 0.5quant 0.975quant
Intercept_L3 0.6598 0.0655     0.5417   0.6561      0.799

Longitudinal outcome (L4, gaussian)
                        mean     sd 0.025quant 0.5quant 0.975quant
Intercept_L4          0.2678 0.0463     0.1770   0.2678     0.3586
year_L4              -0.1474 0.0060    -0.1592  -0.1474    -0.1356
Res. err. (variance)  0.4878 0.0170     0.4555   0.4874     0.5224

Random effects variance-covariance (L4)
               mean    sd 0.025quant 0.5quant 0.975quant
Intercept_L4 0.5012 0.052     0.4068   0.4985     0.6112

Longitudinal outcome (L5, binomial)
                mean     sd 0.025quant 0.5quant 0.975quant
Intercept_L5 -3.7768 0.1841    -4.1377  -3.7768    -3.4159
year_L5       0.2243 0.0324     0.1608   0.2243     0.2878

Random effects variance-covariance (L5)
               mean     sd 0.025quant 0.5quant 0.975quant
Intercept_L5 3.4529 0.6392     2.2997    3.426     4.7976

Longitudinal outcome (L6, binomial)
              mean     sd 0.025quant 0.5quant 0.975quant
Intercept_L6 -1.41 0.1647    -1.7328    -1.41    -1.0872
year_L6       0.13 0.0252     0.0807     0.13     0.1794

Random effects variance-covariance (L6)
              mean     sd 0.025quant 0.5quant 0.975quant
Intercept_L6 5.269 0.9292     3.6123   5.2195      7.251

Longitudinal outcome (L7, gaussian)
                        mean     sd 0.025quant 0.5quant 0.975quant
Intercept_L7         -0.2215 0.0451    -0.3098  -0.2215    -0.1332
year_L7               0.1087 0.0066     0.0957   0.1087     0.1217
Res. err. (variance)  0.6034 0.0213     0.5630   0.6030     0.6465

Random effects variance-covariance (L7)
               mean     sd 0.025quant 0.5quant 0.975quant
Intercept_L7 0.4202 0.0479     0.3333   0.4178     0.5215

Survival outcome (S1)
                              mean     sd 0.025quant 0.5quant 0.975quant
Baseline risk (variance)_S1 0.0299 0.0421     0.0013    0.016     0.1462

Survival outcome (S2)
                            mean     sd 0.025quant 0.5quant 0.975quant
Baseline risk (variance)_S2 0.17 0.2013     0.0125   0.1055     0.7307

Association longitudinal - survival
             mean     sd 0.025quant 0.5quant 0.975quant
CV_L1_S1   1.4001 0.1201     1.1597   1.4014     1.6328
CV_L2_S1  -0.6632 0.1954    -1.0466  -0.6637    -0.2772
CV_L2_S2  -0.5142 0.2752    -1.0439  -0.5183     0.0396
CV_L3_S2   0.6299 0.4178    -0.1868   0.6279     1.4584
SRE_L4_S2 -0.6602 0.3501    -1.3400  -0.6634     0.0385
SRE_L5_S1  0.1403 0.0719    -0.0038   0.1412     0.2792
CV_L6_S2   0.0564 0.1091    -0.1634   0.0581     0.2663
CS_L6_S2  -0.0125 1.1612    -2.3159  -0.0066     2.2564
CS_L7_S1   0.1417 1.1917    -2.1448   0.1218     2.5469

log marginal-likelihood (integration)    log marginal-likelihood (Gaussian) 
                            -94885.27                             -94858.34 

Deviance Information Criterion:  -108646.8
Widely applicable Bayesian information criterion:  -89290.45
Computation time: 116.87 seconds
\end{CodeOutput}
\end{CodeChunk}

\begin{figure}[!ht]
\centering
\includegraphics[scale=0.6]{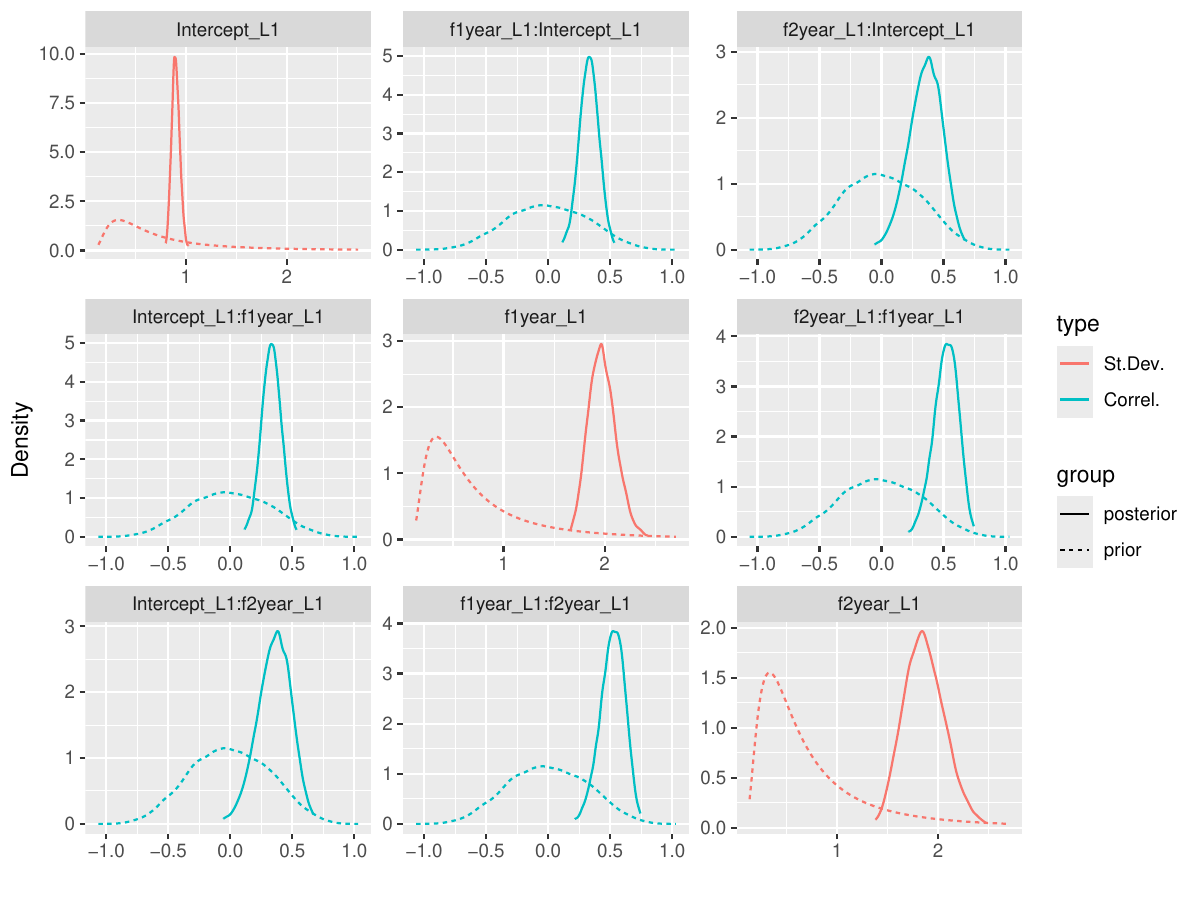}
\caption{Posterior distribution of standard deviation and correlation terms from random effects included in the first longitudinal marker (serum bilirubin) of the model described in Section \ref{Ex2}}
\label{plotSdCor}
\end{figure}

The \code{plot} function returns a series of plots when applied to the fitted model. We only show some of these plots to avoid redundancy as the model includes many outcomes. First, the posterior marginal distributions for fixed effects and residual error terms associated to each longitudinal outcome, for example for the first longitudinal marker, serum bilirubin:
\begin{CodeInput}
pdf("plotL1.pdf", width=6, height=4)
plot(IJ2)$Outcomes$L1
dev.off()
\end{CodeInput}
\begin{figure}[!ht]
\centering
\includegraphics[scale=0.8]{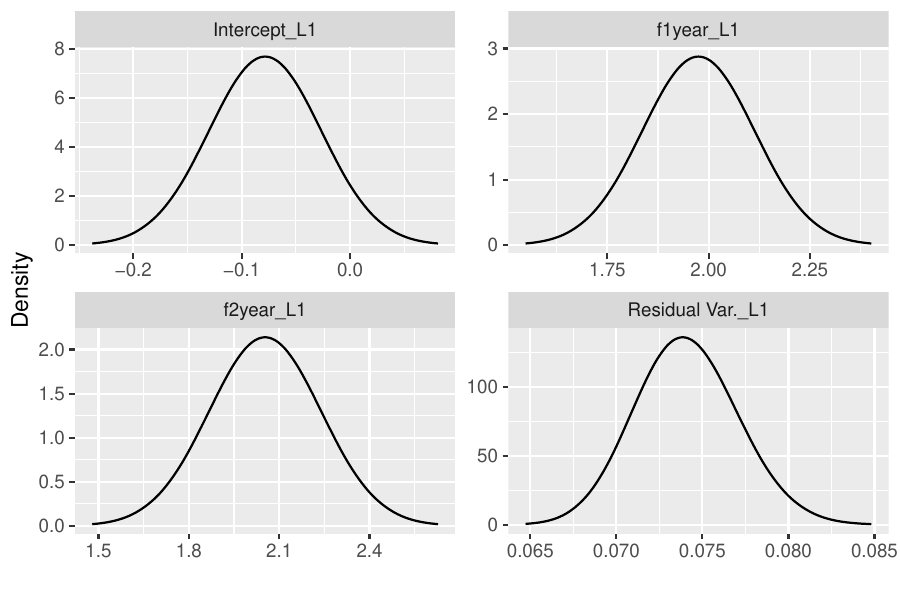}
\caption{Marginal posterior distributions of fixed effects and the residual error variance for the first longitudinal outcome (serum bilirubin) of the model described in Section \ref{Ex2}.}
\label{plotL1}
\end{figure}
The plot is given in Figure \ref{plotL1}. Then, the marginal posterior distribution of all the association parameters:
\begin{CodeInput}
pdf("plotAsso.pdf", width=8, height=6)
plot(IJ2)$Associations
dev.off()
\end{CodeInput}
\begin{figure}[!ht]
\centering
\includegraphics[scale=0.6]{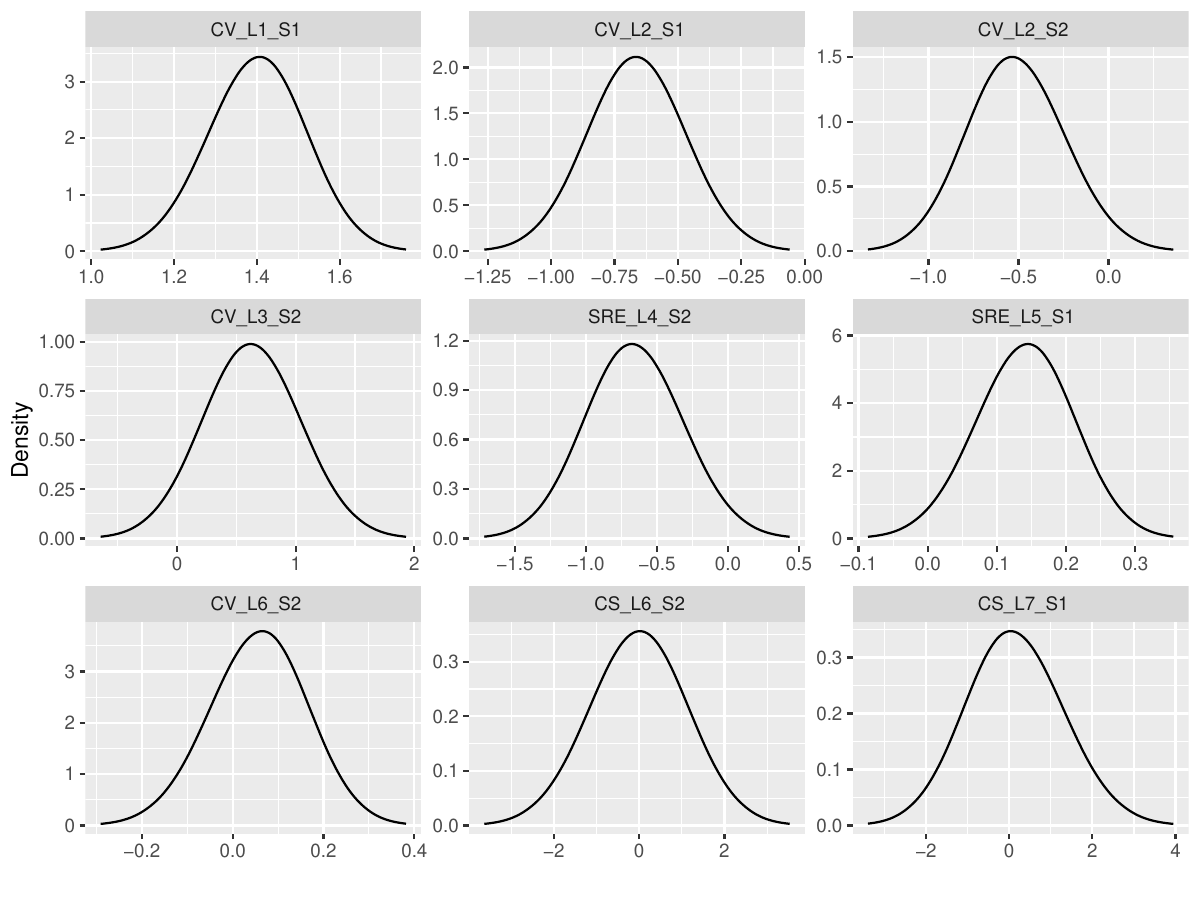}
\caption{Marginal posterior distributions of association parameters of the model described in Section \ref{Ex2}.}
\label{plotAsso}
\end{figure}
The resulting plot is given in Figure \ref{plotAsso}. It is possible to add prior distributions to evaluate the difference between priors and posteriors for all the marginal posteriors, this is useful to identify parameters that may not have enough observed data information and thus will return a posterior similar to the prior, or to evaluate the sensitivity of parameters to different priors specifications:
\begin{CodeInput}
pdf("plotAssoPrior.pdf", width=8, height=6)
plot(IJ2, priors=TRUE)$Associations
dev.off()
\end{CodeInput}
\begin{figure}[!ht]
\centering
\includegraphics[scale=0.6]{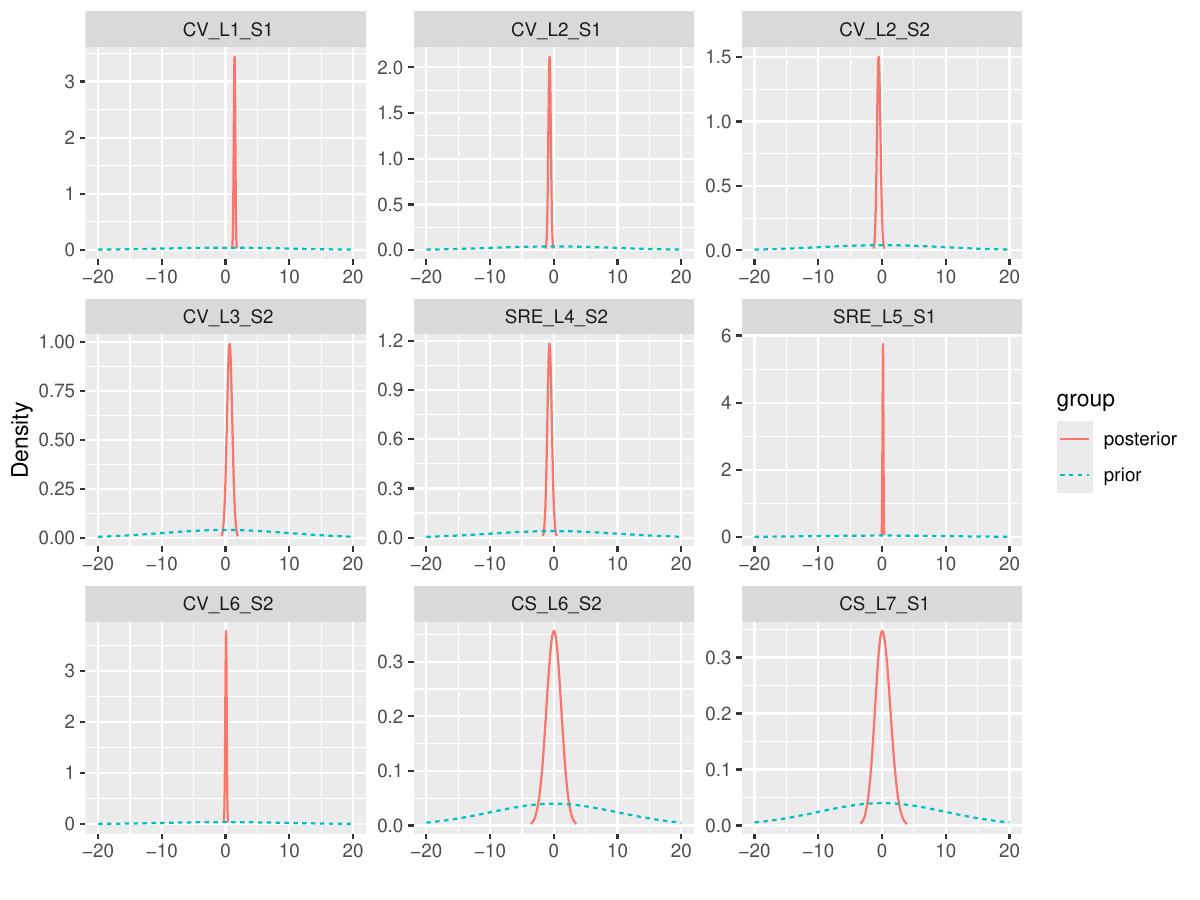}
\caption{Marginal posterior and prior distributions of association parameters of the model described in Section \ref{Ex2}.}
\label{plotAssoPrior}
\end{figure}
In Figure \ref{plotAssoPrior} it is clear that while the default priors are flat, the posteriors are not. We can do the same for the standard deviation and correlation of random effects for the first marker, where the inverse-Wishart priors assume a contraction towards zero for standard deviations and a flat prior over the range of possible correlations with higher density near zero (see Figure \ref{plotSdCor}).
\begin{CodeInput}
pdf("plotSdCor.pdf", width=8, height=6)
plot(IJ2, sdcor=TRUE, priors=TRUE)$Covariances$L1
dev.off()
\end{CodeInput}
Finally, the baseline hazard curves are also part of the plots returned by the \code{plot} function:
\begin{CodeInput}
pdf("plotBaseline.pdf", width=8, height=6)
plot(IJ2)$Baseline
dev.off()
\end{CodeInput}
\begin{figure}[!ht]
\centering
\includegraphics[scale=0.5]{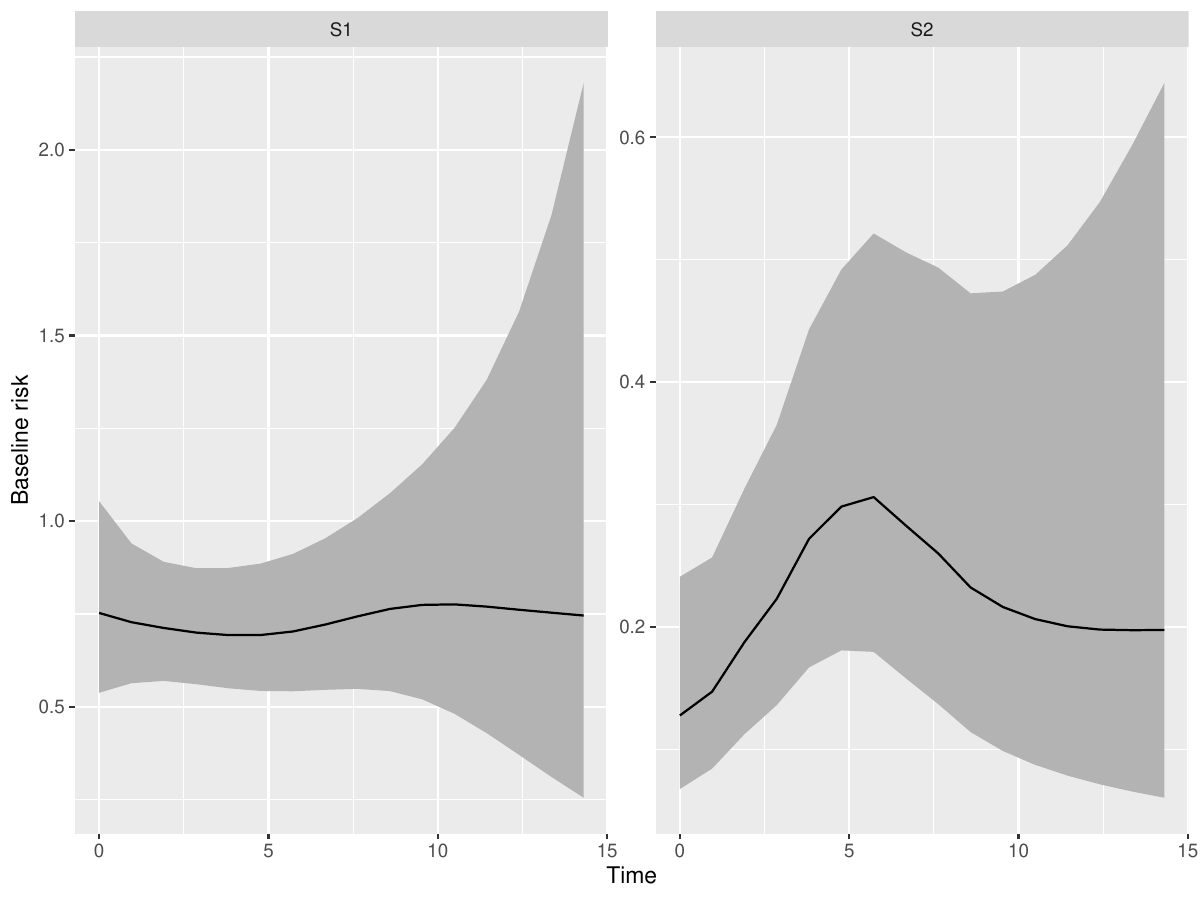}
\caption{Baseline hazard functions of the model described in Section \ref{Ex2}}
\label{plotBaseline}
\end{figure}
Note that when a parametric baseline is assumed (e.g., Weibull), the posterior marginals for the parameters of the baseline distribution are returned, as well as the corresponding curve of the baseline hazard based on the posterior mean, with associated uncertainty.

Now, we illustrate how to use the \code{predict} function. We need to define a new dataset for which we want predictions. Here for simplicity we use data of an individual from the training set but treated as a new individual. In order to illustrate the possibility to deal with dynamic predictions, we define two landmark points, such that the first prediction is based on the first 2 longitudinal measurements while the second prediction is updated to take into account the entire history of longitudinal markers for this individual (i.e., 6 observations). Moreover, we add a line at the end of the new data for predictions where the covariates values are set to 0 (i.e., reference individual) and the outcome values are set to \code{NA}. This will compute predictions of the longitudinal and survival components for the reference individual. By adding a similar line where the variable drug is set to 1 for example, we could compare the average trajectories conditional on drug. This feature can be very useful for inference purposes as it allows to compute the average profiles conditional on covariates values, since it becomes difficult to interpret parameters when complex components are involved such as splines or multiple interactions and it is often easier to visually depict the trajectories of longitudinal and survival components. We set the horizon of the predictions to 14 years.
\begin{CodeChunk}
\begin{CodeInput}
R> NewDat <- pbc2_2[pbc2_2$id==139,]
R> NewDat$id <- 2
R> NewDat <- rbind(NewDat[1:2, ], NewDat, c(3, 0, 0, 0, rep(NA, 7)))
R> NewDat$id[1:2] <- 1
R> HRZ <- 14
R> print(NewDat, digits=1, row.names=F)
\end{CodeInput}
\small
\begin{CodeOutput}
 id drug sex year ascites spiders log_serBilir albumin log_SGOT platelets log_prothrombin
  1    0   1  0.0       0       0        -0.46   -0.16    -0.19       142             0.6
  1    0   1  0.5       0       0        -0.74   -0.36     0.08       120             0.6
  2    0   1  0.0       0       0        -0.46   -0.16    -0.19       142             0.6
  2    0   1  0.5       0       0        -0.74   -0.36     0.08       120             0.6
  2    0   1  1.0       0       0        -0.74    0.02    -0.36       121            -0.1
  2    0   1  2.0       1       0        -0.31   -0.44    -0.51        81             0.6
  2    0   1  2.9       0       0         0.03   -0.08    -0.42        64             1.5
  2    0   1  5.0       0       0         0.81   -1.25    -0.22        59             2.3
  3    0   0  0.0      NA      NA           NA      NA       NA        NA              NA
\end{CodeOutput}
\normalsize
\end{CodeChunk}

We apply the inverse link function in the prediction output by setting argument \code{inv.link=TRUE} so longitudinal predictions are provided in a interpretable scale while properly handling uncertainty. Most link functions are non-linear and change the distribution of Gaussian random-effects on the transformed scale, it is therefore needed to apply the transformation internally on sampled predictions to reflect the uncertainty on the transformed scale. Moreover, it is easier to visualize and interpret the survival curves instead of the hazard curves, in this context it is possible to predict the survival by adding the argument \code{survival=TRUE}. Although here we have competing risks, so we are more interested in the cumulative incidence functions which are computed when setting the argument \code{CIF=TRUE}.

\begin{CodeChunk}
\begin{CodeInput}
R> PRED <- predict(IJ2, NewDat, horizon=HRZ, inv.link=TRUE, CIF=TRUE)
R> sapply(PRED, names)
\end{CodeInput}
\begin{CodeOutput}
$PredL
[1] "id"         "year"       "Outcome"    "Mean"       "Sd"        
[6] "quant0.025" "quant0.5"   "quant0.975"

$PredS
 [1] "id"             "year"           "Outcome"        "Haz_Mean"      
 [5] "Haz_Sd"         "Haz_quant0.025" "Haz_quant0.5"   "Haz_quant0.975"
 [9] "CIF_Mean"       "CIF_Sd"         "CIF_quant0.025" "CIF_quant0.5"  
[13] "CIF_quant0.975"
\end{CodeOutput}
\end{CodeChunk}
The \code{predict} function returns a list of two elements corresponding to predictions for longitudinal and survival outcomes. In the longitudinal part, the first column is the id of the individual from the new dataset and the second column is the measurement time of the prediction (default is 50 equidistant time points between 0 and horizon time). The third column gives the name of the outcome and the other columns are summary statistics of the predictions including mean, sd, median and 95\% credible interval. When the argument \code{inv.link} in the call of the \code{predict} function is set to FALSE, predictions are given for the linear predictors while when set to TRUE, the inverse link is applied and summary statistics are computed on this scale. In the survival part, the same structure is adopted and summary statistics are returned for the value of the hazard function at the same time points as those used in the longitudinal part. The arguments \code{survival} and \code{CIF} adds summary statistics for the survival function and the cumulative incidence function when set to TRUE, respectively. When the argument \code{return.samples} is set to TRUE in the call of the predict function (default is FALSE), summary statistics are replaced by samples (number of samples is \code{Nsample} * \code{NsampleRE}, i.e., number of samples multiplied by number of random effects realizations for each samples). Note that each sample has the same weight here because they are based on sampling from the posterior distribution of the model fitted to the individual data, therefore uncertainty is from marginal parameters and individual random effects deviation (i.e., random effects are sampled conditional on the observed longitudinal for each individual prediction, therefore there is no need to weight the sample with the probability density of predicted curve versus observed longitudinal values as usually done with this type of predictions, which reduces the required number of random effects samples compared to the standard predictions approaches that sample random effects from the population distribution).

Now, we can plot the curves from the \code{PRED} object for each longitudinal outcomes along with CIF for each individual (i.e., the two landmarks and the reference individual).
\begin{CodeInput}
R> plotL <- function(x){
+   PL <- PRED$PredL[PRED$PredL$Outcome==x,]
+    if(x=="log_serBilir"){
+      TITLE <- c("Landmark\n 2 observations", "Landmark\n 6 observations",
+                 "Average for sex = male\n and drug = placebo")
+    }else{
+      TITLE <- c("", "", "")
+    }
+    for(id in unique(NewDat$id)){
+      idn <- ifelse(id==1, x, "")
+      plot(NewDat[NewDat$id==id, c("year", x)], pch=19, xlim=c(0,HRZ),
+           ylim=range(c(na.omit(pbc2_2[, x]), PL[PL$id%in%c(1,2), c(6,8)])), 
+           xlab="", ylab=idn, main=TITLE[id], cex.main=0.9)
+      lines(PL[PL$id==id, c(2,7)], lwd=2)
+      lines(PL[PL$id==id, c(2,6)], lty=2)
+      lines(PL[PL$id==id, c(2,8)], lty=2)
+    }
+  }

R> pdf("Predict.pdf", width=6, height=10)
R> par(mfrow=c(8,3), oma = c(2.5, 1, 1, 1), mai = c(0.2, 0.5, 0.3, 0.05))
R> sapply(unique(PRED$PredL$Outcome), function(x) plotL(x))
R> PS <- PRED$PredS
R> for(id in unique(NewDat$id)){
+    plot(PS[PS$Outcome=="S_1" & PS$id==id, c(2,12)],
+         type="l", lwd=2, xlim=c(0,HRZ), ylim=c(0,1),
+         xlab="Time (Years)", ylab="Event probability")
+    lines(PS[PS$Outcome=="S_1" & PS$id==id, c(2,11)], lty=3)
+    lines(PS[PS$Outcome=="S_1" & PS$id==id, c(2,13)], lty=3)
+    lines(PS[PS$Outcome=="S_2" & PS$id==id, c(2,11)], col=2, lty=2, lwd=2)
+    lines(PS[PS$Outcome=="S_2" & PS$id==id, c(2,13)], col=2, lty=3)
+    lines(PS[PS$Outcome=="S_2" & PS$id==id, c(2,11)], col=2, lty=3)
+    if(id==2){
+      legend("topleft", legend=c("death", "transpl."),
+             lty = c(1,2), lwd=c(2,2), col=c(1,2), bty="n")
+    }
+  }
R> dev.off()
\end{CodeInput}
\begin{figure}[!ht]
\centering
\includegraphics[scale=0.8]{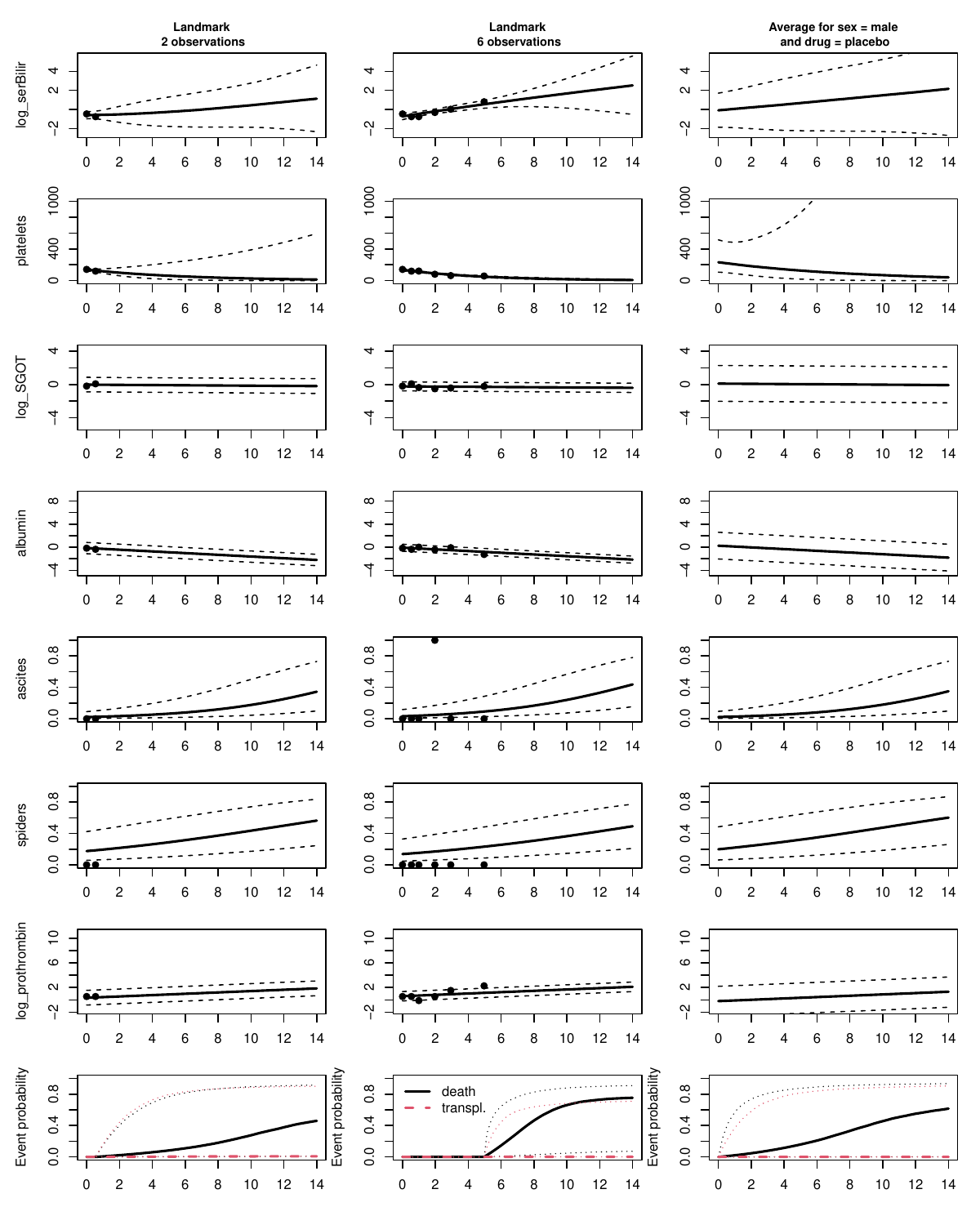}
\caption{Predicted values for each longitudinal marker and cumulative incidence functions.}
\label{Predict}
\end{figure}

The plot is given in Figure \ref{Predict}. The cumulative incidence function that quantifies the probability of each competing event conditional on survival from all events is by default evaluated from the last longitudinal observation time but it is possible to set a starting time for survival predictions by setting the argument \code{Csurv} equal to the desired value. We can see in this plot that when many longitudinal observations are provided (i.e., landmark with 6 observations per longitudinal marker), the uncertainty tends to be reduced compared to predictions with few observations (i.e., landmark with 2 observations per longitudinal markers) where uncertainty is closer to the reference individual without longitudinal information.

\section[Summary]{Summary}

Joint models are useful tools that gain interest in statistics but their development have been limited by the lack of efficient and flexible software. With \pkg{INLAjoint}, we propose a unified framework able to build joint models tailored to specific needs. As opposed to other softwares, there is no restriction in the number of longitudinal and survival outcomes as well as in the structure of the model. This framework is very flexible to build and assemble models together in order to form various joint models. 
 
\pkg{INLAjoint} accommodates various mixed effects regression models for longitudinal data, including generalized linear, proportional odds and zero-inflated mixed effects models that can be combined and linked by correlated random effects. The package supports the fitting of parametric baseline hazards for proportional hazards models, encompassing distributions such as exponential and Weibull. Furthermore, users can approximate a Cox proportional hazards regression model with smooth splines approximation of the baseline hazard. The extension to frailty, mixture cure, competing risks, and multi-state models further broadens the scope of survival modeling within the joint framework. Various association structures can accomodate for the relationship between longitudinal and survival components, including shared random effects, current value, and current slope parametrizations. 

Using the INLA Bayesian algorithm within the \proglang{R} package \pkg{INLA}, our approach mitigates the computational burden associated with iterative estimation techniques in classical software, such as maximum likelihood estimation or Bayesian inference with MCMC sampling. This facilitates the estimation of multivariate joint models with fewer constraints. Fitting such various complex joint models require advanced knowledge and long codes to be fitted directly with the \proglang{R} package \pkg{INLA} and thus motivates the necessity of \pkg{INLAjoint}. The summary of the model fit is more accessible compared to the default output from \pkg{INLA} and allows to easily switch the scale of the parameters, from log hazard ratios to hazard ratios and from variance-covariance to standard deviations and correlations.

Moreover, \pkg{INLAjoint} incorporates several goodness-of-fit criteria for model comparison such as the deviance information criterion, the widely applicable information criterion, and conditional predictive ordinates, allowing to assess the relative performance of fitted models. The inclusion of these metrics facilitates an informed selection process, aiding researchers in choosing models that best capture the underlying dynamics of the data. In response to the practical needs of researchers, \pkg{INLAjoint} includes a user-friendly \code{predict} function that extends across both longitudinal and survival outcomes. This function allows for the generation of predictions for various scenarios, including forecasting future observations, predicting outcomes for new individuals, and imputing missing values as well as for inference purpose to compare expected markers's trajectories conditional on covariates. The \code{plot} function allows to visualize all the parameters's posterior marginals and facilitates prior sensitivity analysis as a critical feature to evaluate the impact of prior specifications on the resulting posterior distributions of model parameters. This analysis aids researchers in assessing the appropriateness of chosen priors and understanding how sensitive the model is to these specifications. By providing insights into the stability and reliability of parameter estimates, prior sensitivity analysis becomes an invaluable tool for model assessment and refinement.

There are limitations to the \pkg{INLAjoint} package. Due to its design for models expressible as latent Gaussian models, some models such as nonlinear mixed effects submodels are excluded. Moreover, while the commonly used association structures between longitudinal and survival submodels are implemented in \pkg{INLAjoint}, the effect of the area under the curve from longitudinal submodels is not available at the moment (while it is available in \pkg{JMbayes2} and \pkg{rstanarm}). There are also limitations in our comparison of \pkg{INLAjoint} with \pkg{JMbayes2} and \pkg{rstanarm} as each method uses different priors, initial values and baseline hazard approximation. However, the purpose of the comparison is to illustrate how we can get the same results as MCMC sampling methods within seconds while the sampling strategy is much slower, such that in this context, the different assumptions are negligible. While in MCMC sampling the approximation error can be reduced by increasing the number of samples, INLA does not rely on sampling and the approximation error only depends on the priors and the likelihood. Since INLA is specifically designed for LGMs and relies on an approximation of the analytical expression of the posteriors, the approximation error is negligible. While MCMC can provide a raw approximation at a low computational cost with few iterations, a high number of iterations is required to match INLA's approximation error.

\pkg{INLAjoint} is a valuable and wieldy tool in the toolkit of practitioners and scientists alike, invoking the powerful INLA methodology while being user-friendly and tailored to longitudinal and survival analysis. It offers promising and exciting possibilities in the realm of joint modeling with the ability to fit very complex models with tiny computational cost. \pkg{INLAjoint} provides a computational framework that can be utilized in the move towards precision and personal medicine by calculating predictions and associated uncertainty in near real-time. We believe that \pkg{INLAjoint} makes joint modeling attainable to all, even those who have complex research questions or a fixed computational budget.

%\section*{Acknowledgments}
\bibliography{refs}
\bibliographystyle{apalike}

\end{document}